\documentclass[
 reprint,
 superscriptaddress,
 amsmath,amssymb,
 aps,
 pr,
prb,
]{revtex4-2}
 \usepackage{graphicx}
\usepackage{dcolumn}
\usepackage{bm}
\usepackage{etoolbox}
\usepackage{amssymb}
\usepackage{amsmath}
\usepackage{physics}
\usepackage{mathtools}
\usepackage{chemformula}
\newcommand{\rom}[1]{\uppercase\expandafter{\romannumeral #1\relax}}
\usepackage{graphicx}  
\usepackage{subfig}
\usepackage{tikz}
 \usepackage{hyperref}
 \hypersetup{
    colorlinks=true,
    citecolor=blue, 
    linkcolor=blue, 
    filecolor=magenta,      
    urlcolor=blue,
}
\begin{document}

\preprint{APS/123-QED}

 \title{Refined Phase Diagram for a Spin-$1$ System Exhibiting a Haldane Phase} 
\author{Mohamad Mousa}
  \email{mmousa@purdue.edu}
 \affiliation{Department of Physics and Astronomy, Purdue University, West Lafayette, Indiana, 47907.}

\author{Birgit Wehefritz–Kaufmann}%
 \email{ebkaufma@purdue.edu}
\affiliation{Department of Physics and Astronomy, Purdue University, West Lafayette, Indiana, 47907.}
\affiliation{Department of Mathematics, Purdue University, West Lafayette, Indiana, 47907.}
\affiliation{PQSEI, Purdue University, West Lafayette, Indiana, 47907.}

 \author{Sabre Kais}%
 \email{skais@ncsu.edu}
 
\affiliation{Department of Chemistry, Purdue University, West Lafayette, Indiana, 47907.}
\affiliation{Department of Physics and Astronomy, Purdue University, West Lafayette, Indiana, 47907.}
\affiliation{PQSEI, Purdue University, West Lafayette, Indiana, 47907.}

\affiliation{Department of Electrical and Computer Engineering, North Carolina State University, Raleigh, NC 27606}

\author{Shawn Cui}%
 \email{cui177@purdue.edu}
\affiliation{Department of Physics and Astronomy, Purdue University, West Lafayette, Indiana, 47907.}
\affiliation{Department of Mathematics, Purdue University, West Lafayette, Indiana, 47907.}
\affiliation{PQSEI, Purdue University, West Lafayette, Indiana, 47907.}

\author{Ralph Kaufmann}%
 \email{rkaufman@math.purdue.edu}
\affiliation{Department of Mathematics, Purdue University, West Lafayette, Indiana, 47907.}
 \affiliation{Department of Physics and Astronomy, Purdue University, West Lafayette, Indiana, 47907.}
 \affiliation{PQSEI, Purdue University, West Lafayette, Indiana, 47907.}

\date{\today} 

 \begin{abstract}
We provide the phase diagram of a 2-parameter spin-1 chain that has a symmetry-protected topological (SPT) Haldane phase using computational algorithms along with tensor-network tools. We improve previous results, showing the existence of a new phase and new triple points. New striking features are the triple end of the Haldane phase and the complexity of phases bordering the Haldane phase in proximity---allowing moving to nearby non-SPT phases via small perturbations. These characteristics make the system, which appears in Rydberg excitons, e.g.\ in Cu$_2$O, a prime candidate for applications.

\end{abstract}
 
\maketitle

\section{Introduction}Topological phases of matter have drawn much attention due to their remarkable properties. Their edge transport protection from disorder and imperfections offers an intrinsic physical immunity to noise. This has found many applications in diverse areas like dissipationless electronics, spintronics, lasers, and quantum computing \cite{Kitaev2001,Sarma2015,Jaworowski2019,bandres2018,He2019}.
 
The modern framework for classifying phases of matter started with the Landau-Ginzburg paradigm, where phases were classified by symmetry breaking \cite{Landau1996}. This picture was challenged by the discovery of the integer quantum Hall effect, the fractional quantum Hall effect \cite{hall80,Laughlin1983,Tsui1982}, and subsequent models such as topological insulators that do not require external magnetic fields \cite{haldane88}, the Haldane phase in 1D spin chains \cite{Haldane1983,Haldane19832}, the quantum spin Hall effect \cite{Kane2005,Kane20052}, string-net models \cite{Levin2005}, and quantum double models \cite{Kitaev2003}. These models exhibit phases that cannot be distinguished locally, can occur at zero temperature, and have a finite energy gap above the ground state. They are collectively called topological quantum phases \cite{Wen2017}.

The now well-established paradigm of quantum phase transitions identifies two phases of gapped ground states if there exists an adiabatic path connecting their Hamiltonians without closing the finite energy gap and avoiding any singularity in the local properties of the ground state \cite{Chen2010}. The topological quantum phases are then classified into three categories: intrinsic topological order which follows this definition without any additional requirements (the topologically nontrivial phase cannot be adiabatically connected to any topologically trivial phase), symmetry-protected topological phases (SPT) which only obey this definition if certain symmetries are imposed on this adiabatic path, and symmetry-enriched topological phases (SET) when the underlying system has intrinsic topological order, and then the phases are enriched by imposing symmetries \cite{Mesaros2013,Chen2010}. 

In 1D, there are no phases with intrinsic topological order; it can be shown that any gapped phase with a local Hamiltonian is adiabatically connected to the trivial phase when no symmetries are imposed \cite{Chen2011}. Consequently, the only topological phases in 1D are SPT phases. SPT phases can have nontrivial edge states. These edge states can implement topological quantum computing similar to Majorana zero modes \cite{Kitaev2001,Sarma2015,Jaworowski2019}.

Haldane mapped the spin-$1$ antiferromagnetic Heisenberg chain exhibiting $SU(2)$ symmetry: ${H}_{\mathrm{Heisenberg}}=J \sum_i  \mathbf{S}_i \cdot  \mathbf{S}_{i+1} \text{ with } J>0$ to the topological quantum field theory of the nonlinear sigma model. He then conjectured that the system would be gapped in 1D chains with integer spin \cite{Haldane19832,Haldane1983}. This was counterintuitive, as the Bethe ansatz for the spin-$\frac{1}{2}$ case is gapless \cite{Bethe1931}. The AKLT model presents an exactly solvable model which retains the Heisenberg model's full $SU(2)$ symmetry: $  H_{\mathrm{AKLT}} =\sum_i\left[\frac{1}{2} \mathbf{S}_i \cdot \mathbf{S}_{i+1}+\frac{1}{6}\left(\mathbf{S}_i \cdot \mathbf{S}_{i+1}\right)^2+\frac{1}{3}\right]$. In the AKLT model, the existence of an energy gap, nontrivial entanglement spectrum, and edge states can be proved analytically \cite{AKLT1987}. It was later shown that the topological Haldane phase could be realized using only the $\mathbb{Z}_2 \times \mathbb{Z}_2$ internal symmetry which is a subset of the $SU(2)$ symmetry \cite{Pollmann2012}. This symmetry is represented by the $\pi$-rotations around the $x$-axis and the $y$-axis: $e^{i \pi S^x}$ and $e^{i \pi S^y}$. Under this symmetry, the Haldane phase belongs to the class of SPT phases \cite{Pollmann2012}.

We start by considering a general spin-$1$ Hamiltonian which has translation and parity invariance in addition to $\mathbb{Z}_2 \times \mathbb{Z}_2$ symmetry (realized by $\pi$-rotations around $x$-axis and $y$-axis), and $U(1)$ rotational symmetry around the z-axis \cite{Klumper1993}. This Hamiltonian has seven constants, $(c_0,\dots, c_6)$, given in Eq.~\eqref{equation:ham}. Many important Hamiltonians, like those of the Heisenberg, AKLT, Ising, and XXZ models are special cases of this Hamiltonian. The model exhibits the topologically nontrivial Haldane phase.  We further include two perturbations $D \left[\left(S_i^z\right)^2-2 / 3\right]$ and $\delta c_1 S_i^z S_{i+1}^z$. These perturbations do not break the symmetries of the Hamiltonian Eq.~\eqref{equation:ham}. 

\begin{equation}\label{equation:ham}
\begin{aligned}
{H}= & \sum_i c_0+(c_1+\delta c_1) S_i^z S_{i+1}^z+c_2\left(S_i^x S_{i+1}^x+S_i^y S_{i+1}^y\right)  \\
  & +c_3\left(S_i^z S_{i+1}^z\right)^2+c_4\left(S_i^x S_{i+1}^x+S_i^y S_{i+1}^y\right)^2 \\
& +c_5\left[S_i^z S_{i+1}^z\left(S_i^x S_{i+1}^x+S_i^y S_{i+1}^y\right) + {h.c.}\right]  \\
 & +c_6\left(S_i^x S_{i+1}^y-S_i^y S_{i+1}^x\right)^2 +D \left[\left(S_i^z\right)^2-2 / 3\right] .
\end{aligned}
\end{equation} 

We study the phase diagram as a function of the parameters $D$ and $\delta c_1$, fixing the parameters $c_0,\dots, c_6$ as in equation \eqref{equation:coeff}, as these parameters are of special interest in applications to Cu$_2$O \cite{Valentin2018}.

Previous studies \cite{Poddubny2019} left some questions open, such as whether all phases meet at triple or higher critical points. They also suggest a direct phase transition from the ferromagnetic to the antiferromagnetic phase. 

We show that all critical points are triple points at most, as shown in Fig.~\ref{fig:phases}. Four of them involve the Haldane phase. Interestingly, there is a new direct phase transition between the Haldane and the anisotropic phase. We also predict the existence of a new phase, which we call the negative D phase, that obstructs a direct phase transition between the ferromagnetic and the antiferromagnetic phases. 

\section{Rydberg Excitons}

The Hamiltonian \eqref{equation:ham} can be realized using a 1D chain of traps of Rydberg excitons in Cu$_2$O \cite{Poddubny2019}. 
The parameters $D$ and $\delta c_1$ then correspond to trap anisotropy and coupling anisotropy in the exciton implementation \cite{Valentin2018,Poddubny2019}. 

Rydberg atoms have an electron excited far from its valence shell, leaving a hole behind. The pair then resembles a Hydrogen atom and exhibits long-range dipole-dipole interactions and robust Rydberg blockade \cite{Urban2009}. Analogously, in a semiconductor environment, an electron can be excited from the filled valence band to the conduction band, leaving a hole behind. This electron-hole pair is called an exciton \cite{Knox1963}. Rydberg atoms and excitons have been studied and used in many applications \cite{Urban2009,Peyronel2012-no,Bernien2017, Bluvstein2023-gr,Taylor2022,sajjan2022}. In these applications, however, they were mostly used as two-level systems distinguishing the ground state from other excited states (qubits). However, our implementation here is different as we are exploiting the angular momentum degree of freedom which offers a richer structure that acts as an intrinsic spin-$1$ system (qutrit). This provides a more direct simulation platform for many-body systems with integer spins. 
\begin{figure}[htbp]
 
\includegraphics[width=0.48\textwidth]{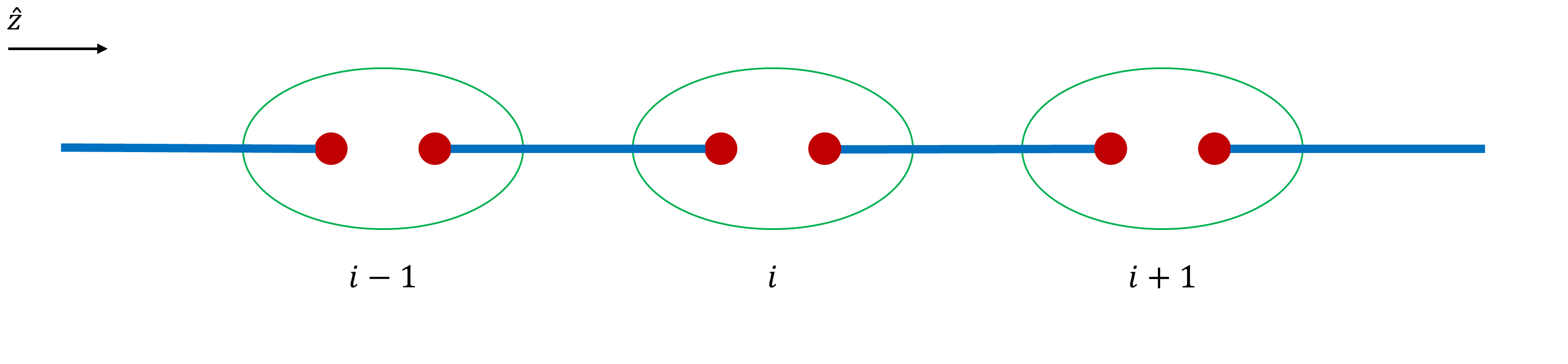}\caption{\label{fig:excitons} The 1D chain of spin-$1$ particles. Blue lines represent the interactions between nearest neighbors. The $z$-axis is aligned with the chain.}
\end{figure}

 The concrete model consists of a 1D chain of trapped excitons which are effectively spin-$1$ particles with nearest-neighbor interactions, see Fig.~\ref{fig:excitons}. The interaction between two neighboring excitons is a Coloumb interaction given by Eq.~\eqref{equation:Vij1}: 
\begin{equation} \label{equation:Vij1}
\begin{aligned}
V^{(i j)}=\frac{e^2}{4 \pi \varepsilon_0 \varepsilon_r} & \left(\frac{1}{\left|\mathbf{r}_e^{(i)}-\mathbf{r}_e^{(j)}\right|}+\frac{1}{\left|\mathbf{r}_h^{(i)}-\mathbf{r}_h^{(j)}\right|}\right. \\
& \left.-\frac{1}{\left|\mathbf{r}_e^{(i)}-\mathbf{r}_h^{(j)}\right|}-\frac{1}{\left|\mathbf{r}_h^{(i)}-\mathbf{r}_e^{(j)}\right|}\right)
\end{aligned}\end{equation}

Here, $\mathbf{r}_e^{(i)}$($\mathbf{r}_h^{(i)}$) denotes the position of the electron(hole) of the atom $i$. This Coulomb interaction can be expanded in terms of the center of mass distance $R_{i j}=\left|\mathbf{R}_i-\mathbf{R}_j\right|$ giving Eq.~\eqref{equation:Vij2}:
\begin{equation} \label{equation:Vij2}
V^{(i j)}=\frac{e^2}{4 \pi \varepsilon_0 \varepsilon_r} \sum_{l, L=1}^{\infty} \frac{\mathcal{V}_{l L}\left(\mathbf{r}_i, \mathbf{r}_j\right)}{R_{i j}{ }^{l+L+1}}
 \end{equation}
 Here, $\mathcal{V}_{l L}\left(\mathbf{r}_i, \mathbf{r}_j\right)$ are the multipole expansion coefficients written as a summation of spherical harmonics $Y_{l m}\left(\hat{\mathbf{r}}_i\right) ,Y_{L-m}\left(\hat{\mathbf{r}}_j\right)$. When the exciton separation  is larger than the Le Roy radius $R_{i j} \gg 2 \cdot\left(\sqrt{\left\langle r_i^2\right\rangle}+\sqrt{\left\langle r_j^2\right\rangle}\right)$, the interaction will be dominated by dipole-dipole interactions ($L=l=1$) \cite{Singer2005,Valentin2018}. Under this condition $\mathcal{V}_{l L}\left(\mathbf{r}_i, \mathbf{r}_j\right)$ will simplify to Eq.~\eqref{equation:Vij3}:
 \begin{equation} \label{equation:Vij3}
V^{(i j)} \approx \frac{e^2}{4 \pi \varepsilon_0 \varepsilon_r}\left(\frac{r_i r_j}{R_{i j}^3}-\frac{3\left(\mathbf{r}_i \cdot \mathbf{R}_{i j}\right)\left(\mathbf{r}_j \cdot \mathbf{R}_{i j}\right)}{R_{i j}^5}\right)
\end{equation}
The resulting Hamiltonian for the pair of excitons can be expanded in terms of their wave functions $\psi_{n_i, l_i, m_i}\left(\mathbf{r}_i\right)$ and $\psi_{n_j, l_j, m_j}\left(\mathbf{r}_j\right)$ for $l_i=l_j=1$. After exact diagonalization, the resulting eigenvectors were found to not depend strongly on $n$ for $15 \leq n \leq 25$ \cite{Valentin2018}. The Hamiltonian corresponding to the eigenfunctions can then be written in terms of the seven spin-$1$ operator terms in Eq.~\eqref{equation:ham} \cite{Poddubny2019}. The eigenvalues will fix the seven constants $c_0,\dots, c_6$ as presented in Eq.~\ref{equation:coeff}.

\begin{equation}\label{equation:coeff}
\begin{array}{r@{}l r@{}l r@{}l r@{}l}
c_0 & = -5.58 \epsilon_0, & c_1 & = 9.53 \epsilon_0, & c_2 & = -8.97 \epsilon_0, & c_3 & = 1.27 \epsilon_0, \\
c_4 & = 6.59 \epsilon_0, & c_5 & = -3.18 \epsilon_0, & c_6 & = 5.04 \epsilon_0. & & 
\end{array}
\end{equation}
Here, $\epsilon_0 = 10^{-4} n^{11} \hbar \mathrm{s}^{-1} \times \mu \mathrm{m}^6 / R^6$ where $n$ is the quantum number which ranges from 15 to 25, and $R$ is the distance between traps \cite{Valentin2018}.
  
\section{Order Parameters}\label{other order}
\begin{figure}[htbp]
\centering
\subfloat[The previous phase diagram calculated in \cite{Poddubny2019}.\label{fig:phasea}]{\includegraphics[width=0.4\textwidth]{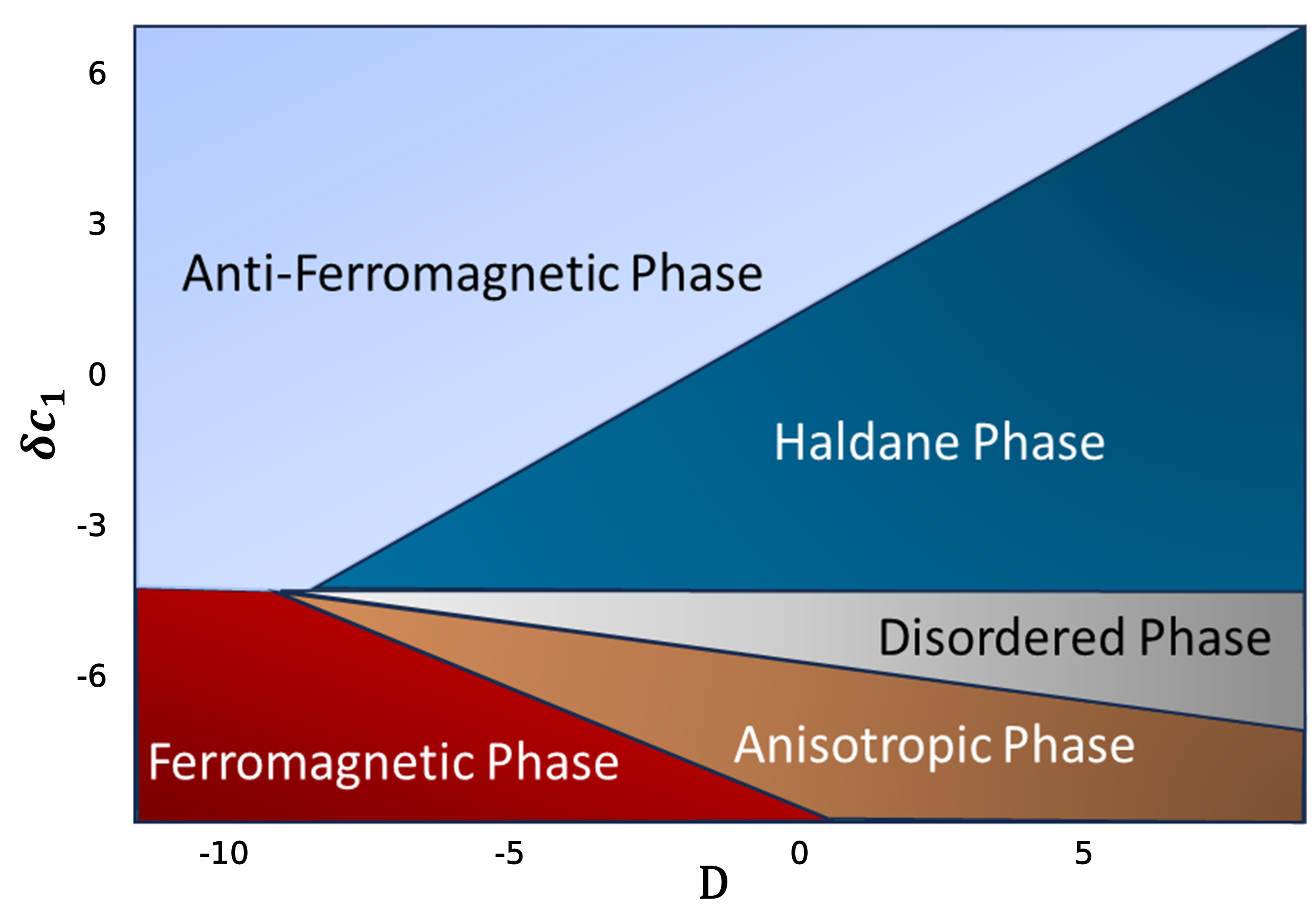}} \hfill
\subfloat[The updated phase diagram displays six triple points $\left(A-F\right)$ and the new negative D phase. Line \rom{5} is used to show the existence of point 
$E$.  \label{fig:phaseb}]{\includegraphics[width=0.4\textwidth]{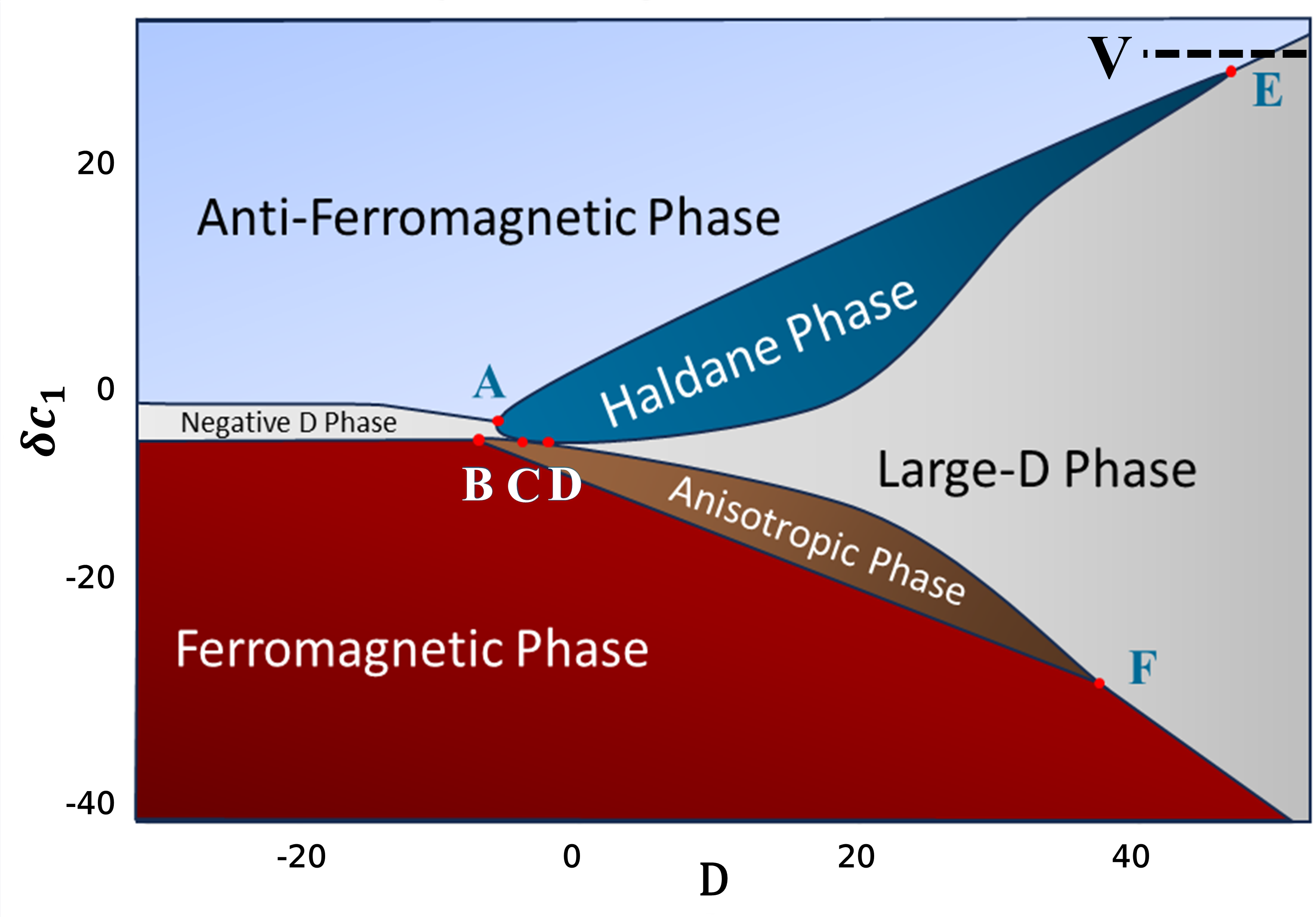}}\hfill  
\subfloat[Schematic figure for the lines used to prove the topology of the critical points in the central region. Accurate line positions are given in the discussion.  \label{fig:phasec} \label{fig:XY1a}] {\includegraphics[width=0.4\textwidth]{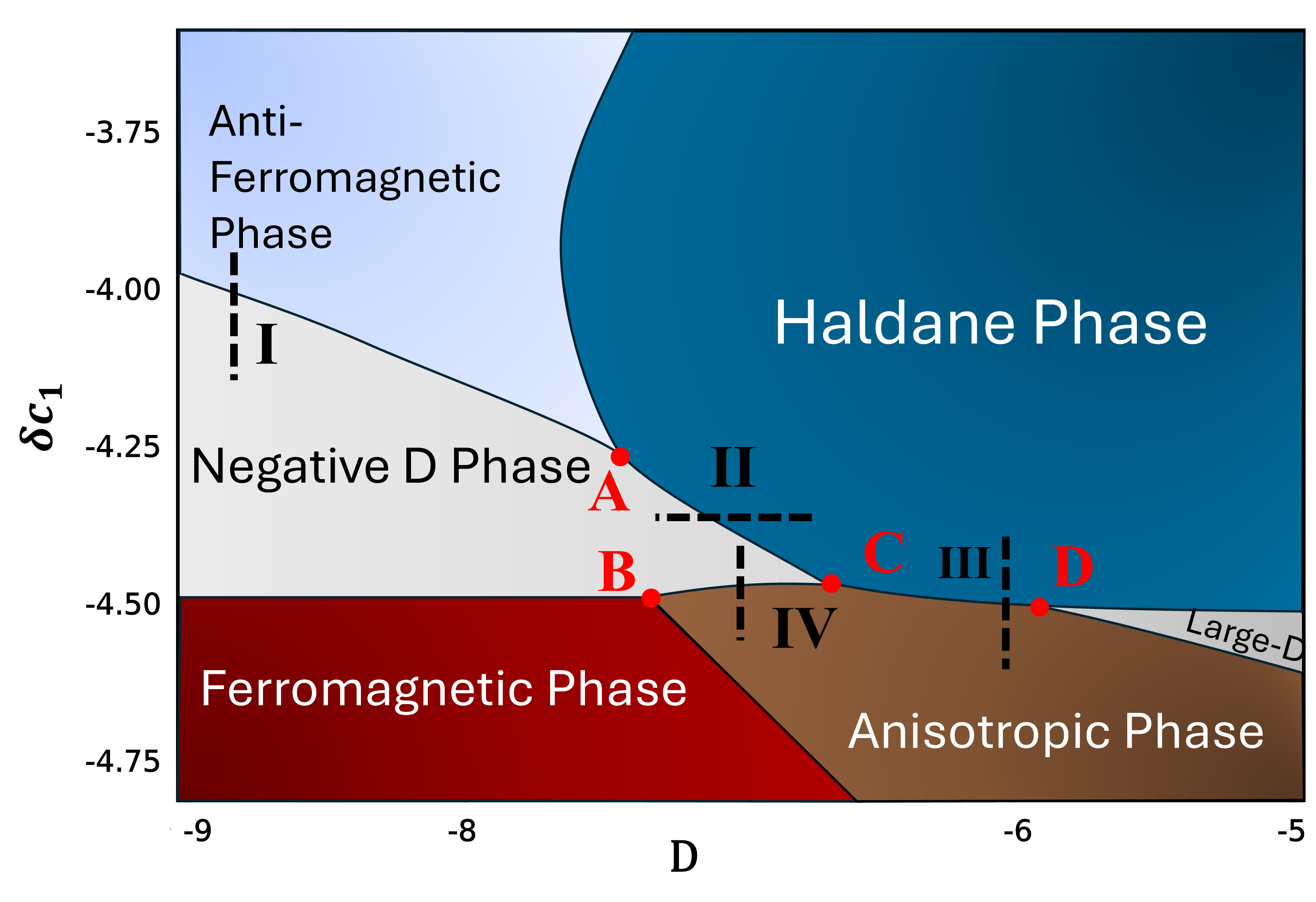}} 
\caption{  Comparison of the old phase diagram (a), the updated phase diagram (b), and the close-up phase diagram (c) for the Hamiltonian Eq.~\eqref{equation:ham} for different values of trap anisotropy $D$ and coupling anisotropy $\delta c_1$.} \label{fig:phases}
\end{figure}

The model exhibits six phases: antiferromagnetic, ferromagnetic, anisotropic, Haldane, negative D, and large D, as shown in Fig.~\ref{fig:phases}. While the anisotropic phase has algebraically decaying spin correlations in the XY plane, 
the negative D and large D phases have exponentially decaying spin correlations in all directions.

For the magnetic phases, we use the antiferromagnetic order parameter:
 $ {O}_{\text{AFM}} = \lim_{|i-j| \to \infty} (-1)^{|i-j|} \langle S_i^z S_{j}^z  \rangle$ and the ferromagnetic order parameter which is just the magnetization in the $z$-axis:
$O_{S_z} =  \frac{\langle \sum_i S_i^z  \rangle}{N}$. The anisotropic phase has some z magnetization and algebraically decaying spin correlations in the x- and y-directions. We can detect it using the spin correlation function of x or y for distant spins \begin{equation}\label{equation:sxsx} {O}_{\text{SxSx}} = \langle S^x_{N/3} S^x_{2N/3} \rangle   \end{equation} For the Haldane phase, we use the string order parameter \cite{dennijs1989}. More details on the string order parameter and its generalization are provided in appendix \ref{subsection:genstring}.:
\begin{equation}\label{equation:string} {O}_{\text{string}} = \lim_{|i-j| \to \infty} \langle S_i^{z}  [ \Pi_{k=i+1}^{j-1} e^{i \pi S_k^{z}}  ] S_j^{z} \rangle \end{equation} It should be noted that the string order parameter will necessarily have a zero value for any trivial SPT phase if the phase is symmetric. The string order parameter can have a non-zero value for a trivial SPT phase if the symmetry is broken. For example, the N\'eel state: $\psi _{\text{ N\'eel}} = \ket{1}\otimes \ket{-1}\otimes \ket{1}\otimes \ket{-1}\otimes \cdots$ will have a non-zero value for the string order parameter because it breaks the  $\mathbb{Z}_2 \times \mathbb{Z}_2$ symmetry, namely the $e^{i \pi S_x}$ symmetry. This issue is resolved by first inspecting the symmetry-breaking phases and then classifying the symmetry-preserving phases using the topological order parameters. In our case, it is also convenient to combine order parameters to distinguish the Haldane phase from other topologically trivial phases using only one order parameter. We define:

\begin{equation}\label{equation:adjusted}
{O}_{\text{Haldane}} =  \lim_{|i-j| \to \infty} \langle \hat{O}_{\text{string}}  + (-1)^{|i-j|}   S_i^z S_j^z  \rangle
\end{equation}

 This new order parameter will be non-zero in the Haldane phase and will be zero in the antiferromagnetic phase. The new negative D phase has exponentially decaying correlations in the XY plane, unlike the anisotropic phase. It is characterized by the absence of all order parameters except ${O}_{S_z^{2}}$:\begin{equation}\label{equation:sz2}  {O}_{S_z^{2}} = \langle \left(S^z\right)^2 \rangle \end{equation} The large D phase is distinguished by the vanishing of all order parameters except $\hat{O}_{S_z^{2}}$, which asymptotically reaches zero. These order parameters are then used to construct the full phase diagram Fig.~\ref{fig:phases} as summarized in Table~\ref{tab:ordersandphases}.

 \begin{table}[h]
 \begin{ruledtabular}
 \begin{tabular}{llc} 

\textrm{Phase}& 
\textrm{Non-zero Order Parameter(s)}& 
\textrm{SPT}\\
 \colrule
Haldane & $  {O}_{\text{String}}$, $  {O}_{\text{Haldane}}$ and $ {O}_{S_z^{2}}$ & Nontrivial \\
Negative D & $ {O}_{S_z^{2}}$ & Trivial \\
Anisotropic&  $ {O}_{\text{SxSx}}$,$ {O}_{S_z}$ and $ {O}_{S_z^{2}}$ & Trivial \\
Antiferromagnetic & $ {O}_{\text{AFM}}$, $ {O}_{S_z^{2}}$ and $  {O}_{\text{String}}$& Trivial \\
Ferromagnetic & $ {O}_{S_z}$ and $ {O}_{S_z^{2}}$ & Trivial \\
Large D & $ {O}_{S_z^{2}} < 1$   & Trivial \\
 
\end{tabular}
\end{ruledtabular}

\caption{Order parameters identifying different phases}
\label{tab:ordersandphases}
\end{table}

 \section{Computational Parameters} We used the density matrix renormalization group (DMRG) algorithm to obtain the extended phase diagram for the finite system and the variational uniform matrix product states (VUMPS) algorithm to simulate the system in the thermodynamic limit \cite{White1992,itensor,VUMPS}. Both algorithms use the matrix-product state (MPS) representation of the wave function \cite{White1992,Fannes1992,Garcia2007}: \begin{equation}
\ket{\psi} = \Tr ( A^{p_1} A^{p_2} \cdots A^{p_N} ) \ket{p_1} \otimes \ket{p_2} \dots  \otimes \ket{p_N} 
\end{equation}

This is a representation of a periodic and translationally invariant state. The indices $p_i$ run over the spin-$1$ degrees of freedom. The matrices $A^{p_i}$ have dimensions $\chi \times \chi$, where $\chi$ is the bond dimension, and it captures entanglement between neighboring sites. The product state, for example, is represented by $\chi=1$; see appendix \ref{mpsconst}. for more details. In DMRG calculations, we used bond dimensions of up to $\chi = 300$. In VUMPS calculations, we used a two-site unit cell and bond dimensions of up to $\chi = 300$. With this choice, we already achieved energy convergence of better than $0.01\%$ in both algorithms.

\section{Results}The phase diagram of the model was first considered in \cite{Poddubny2019}, as presented in Fig.~\ref{fig:phasea}. We obtained an updated phase diagram shown in Fig.~\ref{fig:phaseb},and a close-up of the central region is shown in Fig.~\ref{fig:phasec}. There are six triple points, $A-F$. Four of them are in contact with the Haldane phase. A new phase, which we call the negative D phase, appears in a narrow strip between the ferromagnetic and antiferromagnetic phases and extends as $D\to - \infty$.

\begin{figure}[htbp]\subfloat[Heat map of the Haldane order parameter Eq.~\eqref{equation:adjusted} which is only non-zero in the Haldane phase. \label{fig:ordera}]{\includegraphics[width=0.48\textwidth]{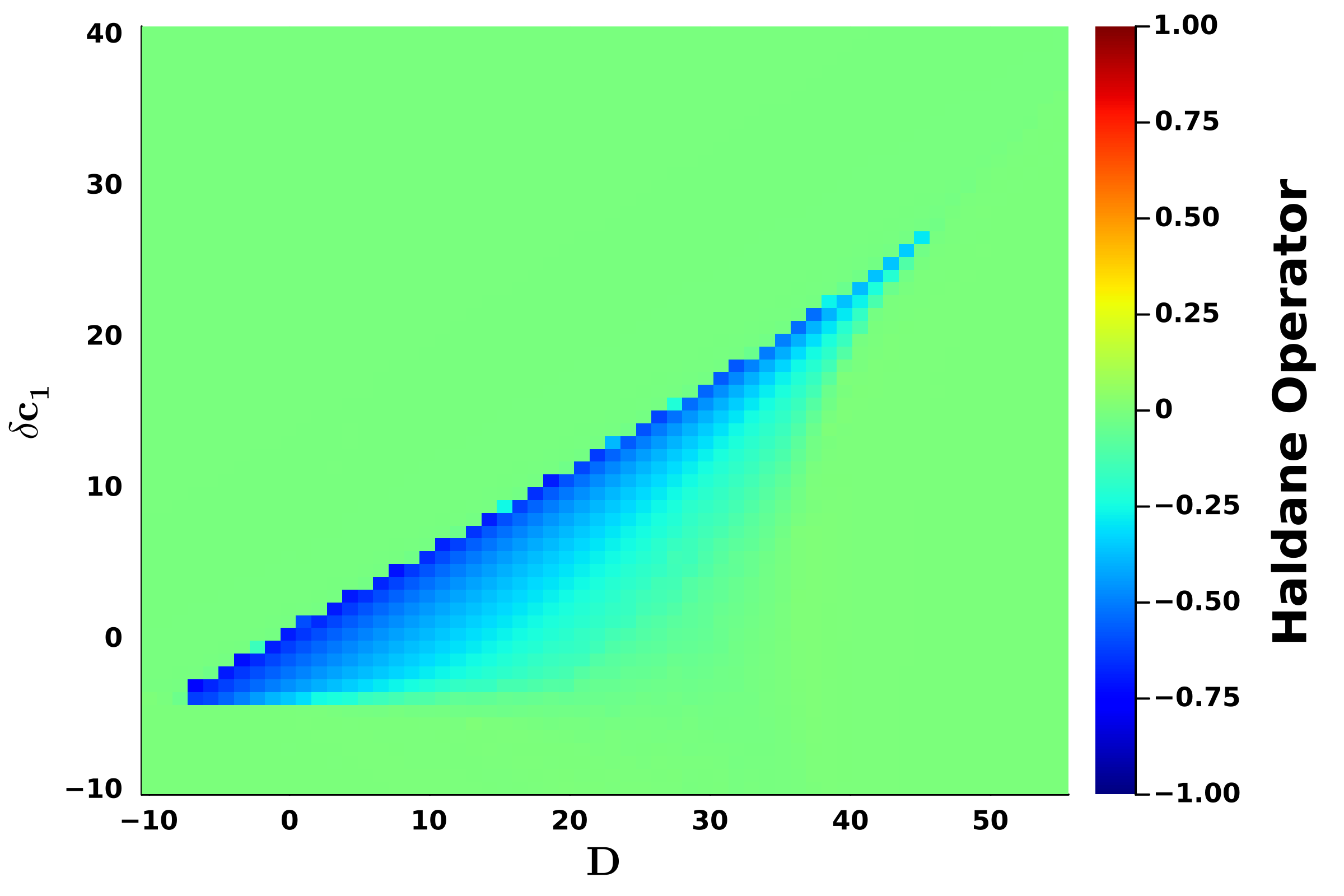}} \hfill
 
\subfloat[Heat map of the sum of string order parameter ($-O_\text{string}$), ferromagnetic order parameter ($O_{S_z}$) and antiferromagnetic order parameter (-${O}_{\text{AFM}}$). \label{fig:orderb}]{\includegraphics[width=0.48\textwidth]{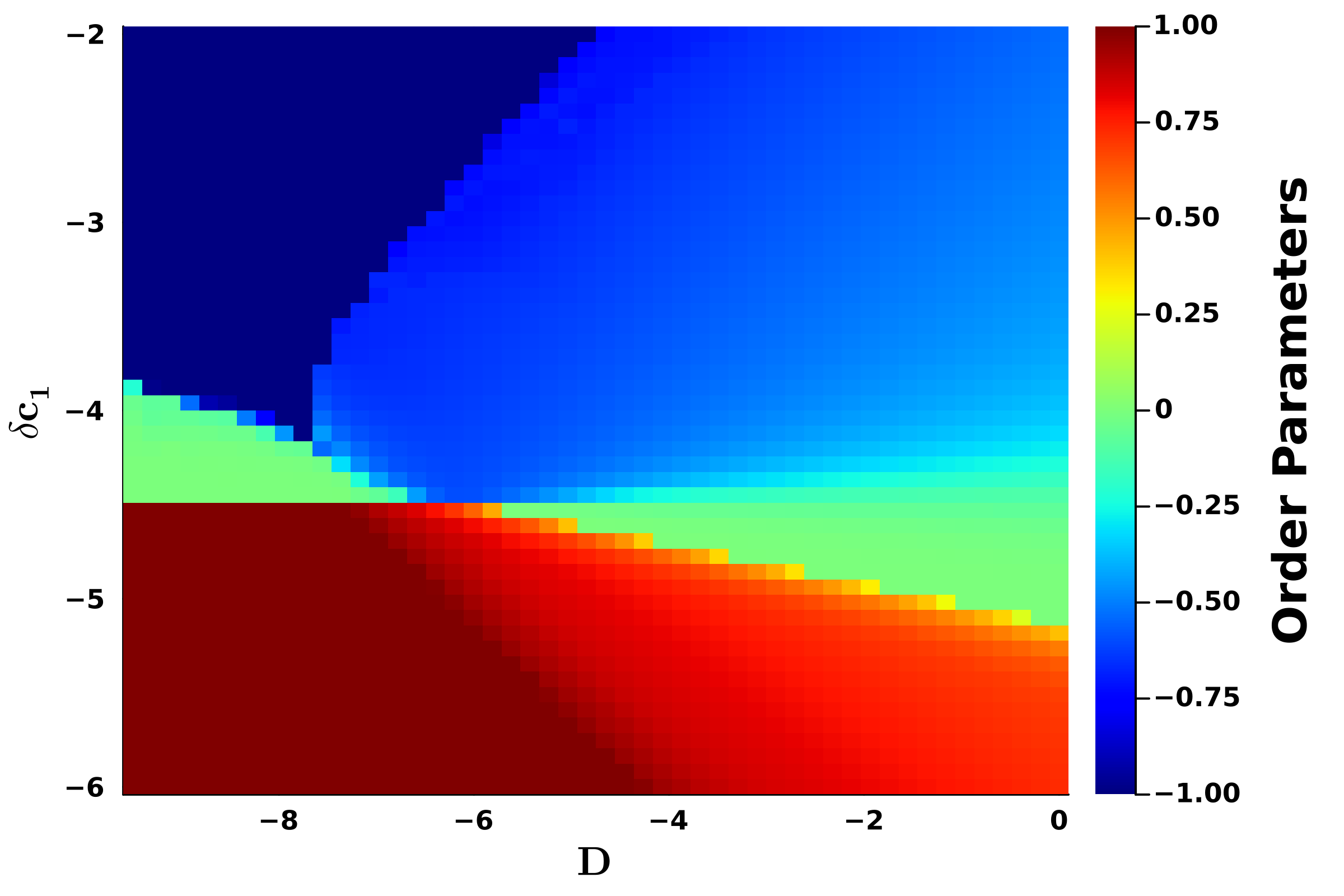}} \hfill
\caption{ The numerical data for the phase diagram of the system in the thermodynamic limit obtained through the VUMPS algorithm.} \label{fig:numphases}
\end{figure}

 Two important numerical results of the system in the thermodynamic limit, using the VUMPS algorithm \cite{VUMPS}, are shown in Fig.~\ref{fig:numphases}.   In Fig.~\ref{fig:ordera}, the Haldane order parameter $ {O}_{\text{Haldane}}$ defined in Eq.(\ref{equation:adjusted}) highlights the region hosting the topological Haldane phase. In Fig.~\ref{fig:orderb}, we added the Haldane order parameter and the ferromagnetic order parameter ${O}_{{S_z}}$ and $ {O}_{\text{AFM}}$ to visualize the phases and critical points in the central region. Fig.~\ref{fig:orderb} is then used to synthesize Fig.~\ref{fig:phasec}.  

 \subsection{Central Region}
\begin{figure}[htbp]\subfloat[ Line \rom{2} in Fig.~\ref{fig:phasec}. The direct phase transition from the negative D phase to the Haldane phase at $\delta c1 = -4.35$ for $N=75$ and open boundary conditions (OBC). \label{fig:XY1b}]{\includegraphics[width=0.48\textwidth]{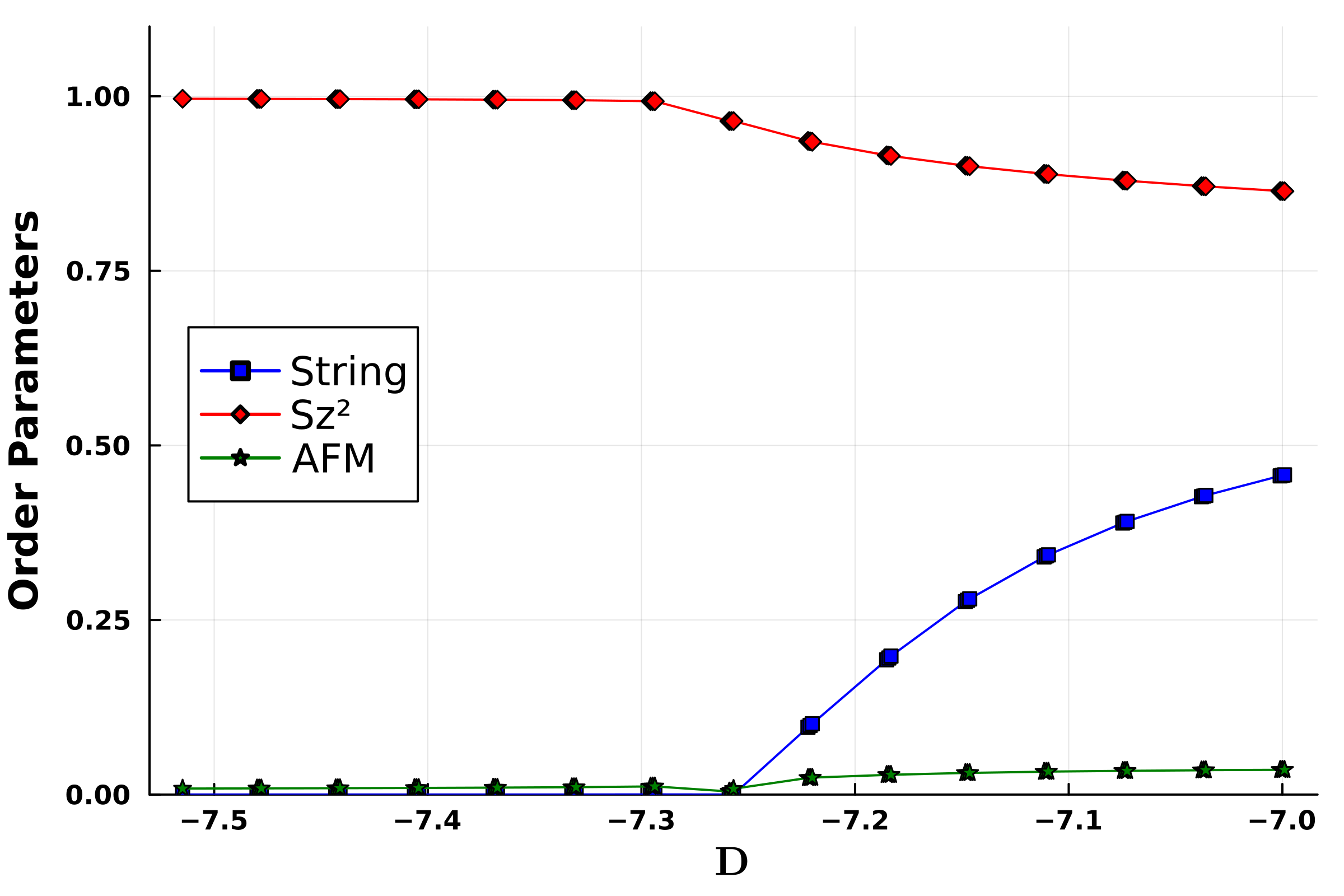}} \hfill\subfloat[ Line \rom{3} in Fig.~\ref{fig:phasec}. The direct transition from the anisotropic phase to the Haldane phase at $D=-6.2$ for $N=60$ and OBC.  \label{fig:XY1d}]{\includegraphics[width=0.48\textwidth]{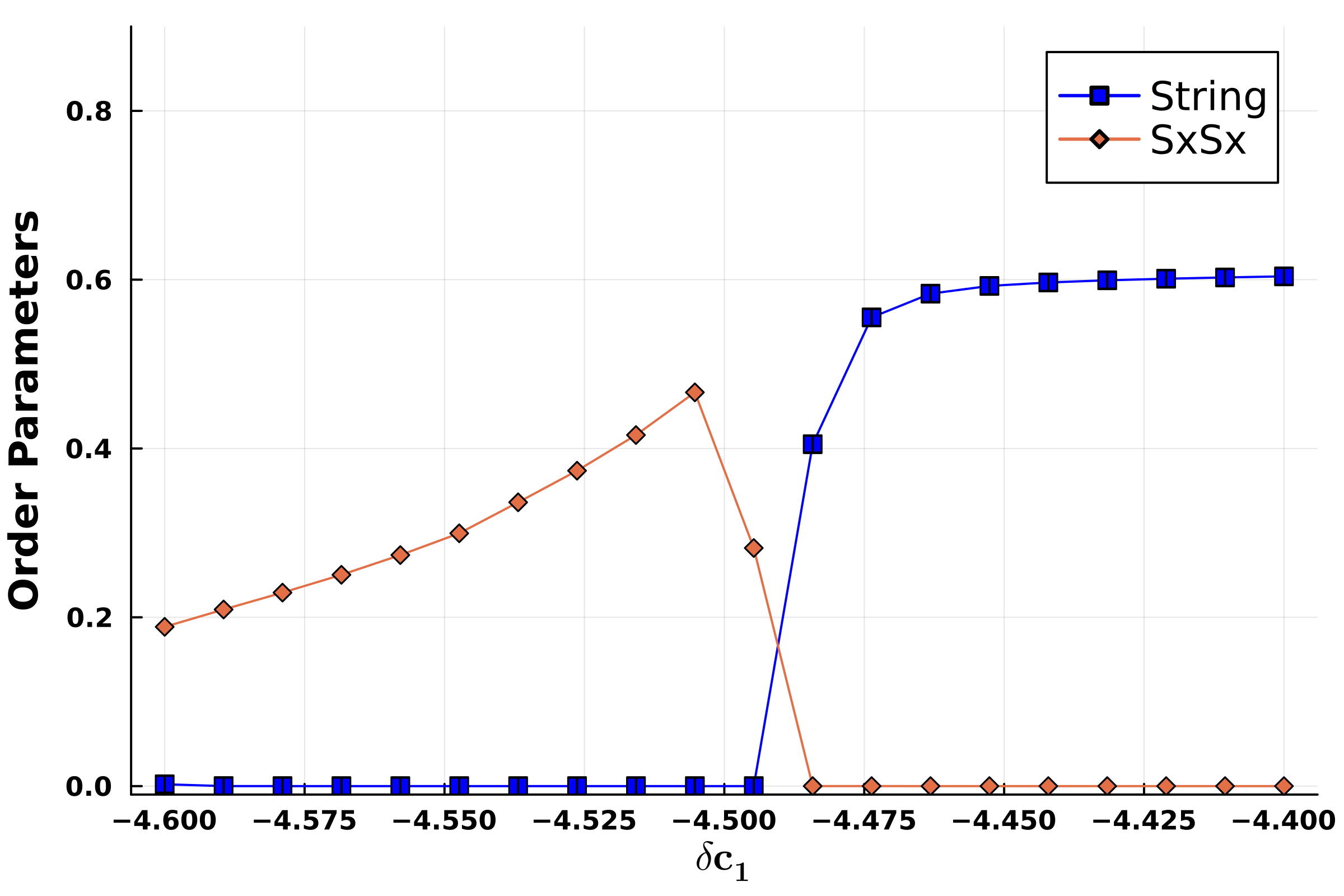}} \hfill\caption{Proving the existence of critical points $A-D$. Error bars are less than the marker size for both figures. }\label{fig:XY1}
\end{figure}

The existence of the triple points $A, B, C$, and $D$ can be proven by investigating direct phase transitions before and after the triple points. This is achieved by studying five paths in the phase diagram, as shown in Fig.~\ref{fig:XY1a}. We show two of these lines as examples; others will be treated in the finite-size scaling analysis below. The data were simulated using the DMRG algorithm with a maximum bond dimension of $\chi = 200$. Line \rom{2} is presented in Fig.~\ref{fig:XY1b}, where a direct transition from the negative D phase to the Haldane phase occurs. The vanishing of $ {O}_{\text{AFM}}$ throughout this path shows the termination of the antiferromagnetic phase after point $A$. The negative D phase has $ {O}_{S_z^2} \approx 1$, which starts to decrease after the phase transition. Fig.~\ref{fig:XY1d} follows line \rom{3} and shows a direct transition from the anisotropic phase, with algebraically decaying spin correlations in the XY plane, to the Haldane phase, with exponentially decaying spin correlations, proving the existence of the triple point $D$.  

\subsection{Asymptotic Behavior} 
\begin{figure}[htbp]
 
\includegraphics[width=0.48\textwidth]{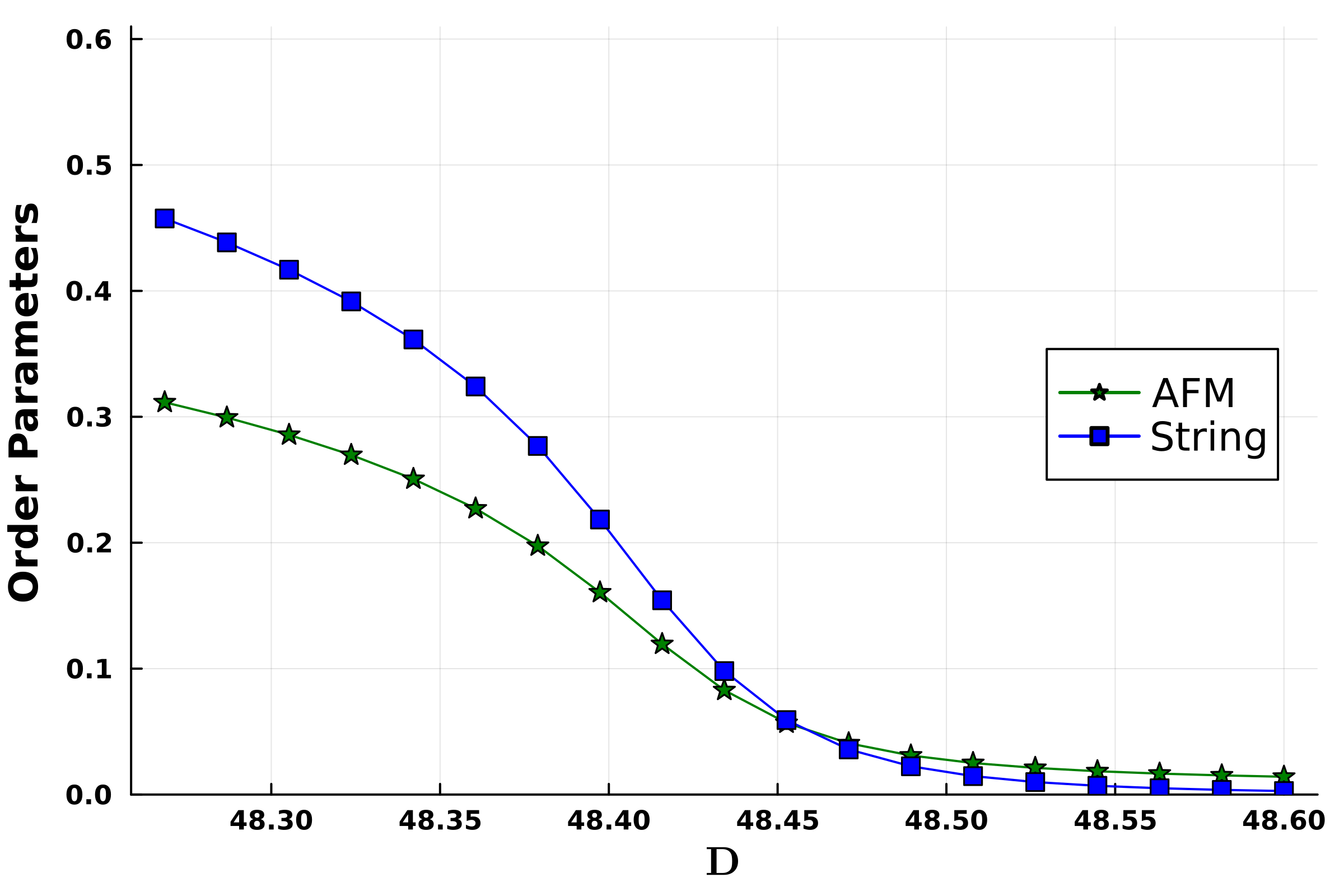}
\caption{\label{fig:afterE} Line \rom{5} in Fig.~\ref{fig:phaseb}. A direct phase transition from the antiferromagnetic to the large D phase at $\delta c_1 = 30$. The simultaneous vanishing of the two order parameters shows the termination of the Haldane phase after point E. Data shown are for $N = 75$ and OBC. Error bars are smaller than marker size.  }
\end{figure}

Upon extending the parameter space we discover that the Haldane and the anisotropic phases end with triple points $E$ and $F$, as shown in Fig.~\ref{fig:phaseb}. This is expected from the asymptotic behavior of the two parameters $D$ and $\delta c_1$. When $(D,\delta c_1) \to (\infty,\infty)$, the model resembles an Ising Hamiltonian with uniaxial anisotropy $H \approx \sum_i c_z S_i^z S_{i+1}^z +D \left(S_i^z\right)^2$. In this regime, we expect only antiferromagnetic or large D phases. Similarly, when $(D,\delta c_1) \to (\infty,-\infty)$, only ferromagnetic or large D phases will survive. The existence of the triple point $E$ is evidenced by a direct phase transition between the antiferromagnetic and large D phases at $\delta c_1 = 30$ (Line \rom{5}), proving that the Haldane phase ends, see  Fig.~\ref{fig:afterE}. Analogously, the existence of the triple point $F$ is proven by a direct transition from the ferromagnetic phase to the large D phase (not shown here).

\section{Finite-size Analysis} \label{sec:fsss }  

\subsection{Antiferromagentic-negative-D  transition  (Line \rom{1})}\label{sec:afd}
  
\begin{figure}[htbp]
 
\includegraphics[width=0.48\textwidth]{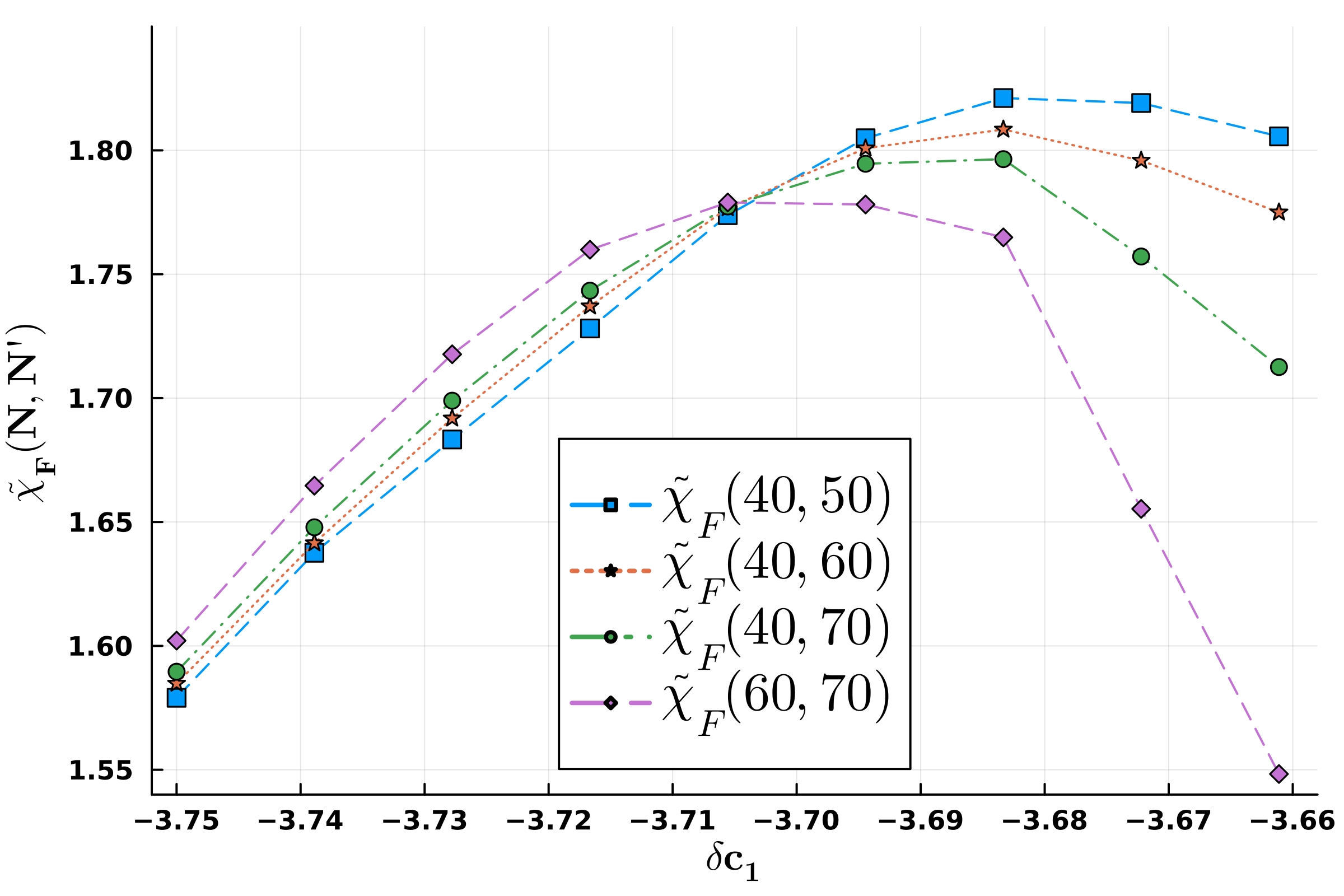}

\caption{\label{fig:fssdaf} Reduced fidelity susceptibility ($\tilde{\chi}_F$) for different system sizes as a function of $\delta c_1$ at fixed $D=-10$. This is the transition from the antiferromagnetic to the negative D phase along line \rom{1} in Fig.~\ref{fig:XY1a}. The functions collapse into one point at the critical value of $\delta c_1$.}
\end{figure}

We perform finite-size scaling analysis to show that the new phases and triple points survive the thermodynamic limit. We start with the phase transition from the antiferromagnetic phase to the negative D phase at $D = -10$. This transition is continuous and according to finite-size scaling theory \cite{Fisher1972,Ren2020}, we can use fidelity susceptibility to determine the position of the critical point in the thermodynamic limit. Fidelity susceptibility is defined as:     
$  \chi_F=    \langle \frac{\partial}{\partial \lambda} \psi  \rvert \frac{\partial}{\partial \lambda} \psi \rangle  - \langle \frac{\partial}{\partial \lambda} \psi \rvert \psi \rangle \langle\psi \rvert \frac{\partial}{\partial \lambda} \psi \rangle$ where $\lambda$ is the controlling parameter. The value of the fidelity susceptibility at the critical point is expected to scale as a power law in the system size $\chi_F(\lambda_c) \propto N^{\mu -1}$ \cite{Ren2020,GU2010,Kais2003}.  We can then define a function of  $\delta c_1$:
\begin{equation}\label{equation:chi}
\tilde{\chi}_{F}(\delta c_1,N,N^{\prime}) \coloneqq \frac{\log(\chi_F(\delta c_1,N))-\log(\chi_F(\delta c_1,N^{\prime}))}{\log(N)-\log(N^\prime)}\end{equation}
The function $\tilde{\chi}_{F}$ should be independent of the system size at the critical value of $\delta c_1$. The functions $\tilde{\chi}_{F}$ versus $\delta c_1$ for different $N$ and $N^{\prime}$ were simulated using the DMRG algorithm for system sizes $N \in \{ 40,50,60,70\}$ and with maximum bond dimension $\chi = 200$. We show that the functions collapse at one point in Fig.~\ref{fig:fssdaf}. The lines intersect near the point $\delta c_1 = -3.70(5)$.  

\subsection{Negative-D-Haldane transition (Line \rom{2})}
  
\begin{figure}[htbp]
 
\includegraphics[width=0.48\textwidth]{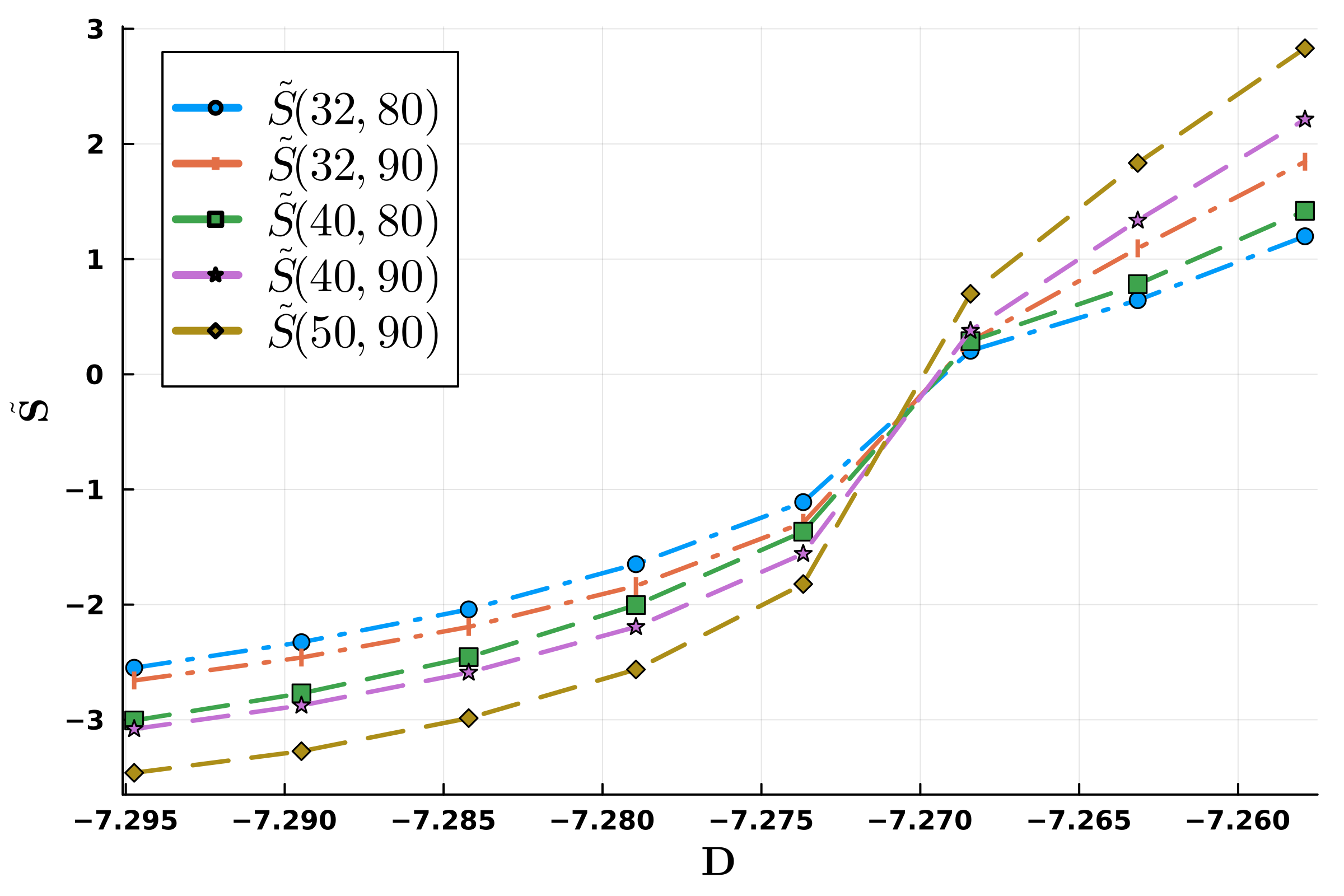}

\caption{\label{fig:fssdhaldane} Reduced string order parameter ($\tilde{S}$) defined in Eq.~\ref{equation:stilde} as a function of $D$ for different system sizes at $\delta c_1 = -4.35$. It shows Line \rom{2} in Fig.~\ref{fig:XY1a} which tracks the transition from the negative D to the Haldane phase. The functions collapse into one point at the critical value of $D$.  }
\end{figure}

We use the non-local string order parameter as a signature for the topological Haldane phase since it vanishes in the negative D phase as seen in Table~\ref{tab:ordersandphases}. The finite-size scaling of the string order parameter is expected to show a power-law dependence near the critical point for continuous transitions \cite{Tonooka2007,Ueda2008}: 
\begin{equation}
 {O}_{\mathrm{string}}^\alpha\left(N, D \right)=N^{-\eta_\alpha} f\left(\left(D_{\mathrm{c}}-D\right) N^{1 / \nu_\alpha}\right) 
 \end{equation}
Where $f$ is a universal function, $\eta_\alpha$ and $\nu_\alpha$ are critical exponents, and the index $\alpha$ represents the direction of the string order parameter ($z$ in our case). We can then define a size-independent function analogously to Eq.~\ref{equation:chi}:
\begin{equation}\label{equation:stilde} \tilde{S}(D,N,N^{\prime}) \coloneqq \frac{\log( {O}_{String}(D,N) /  {O}_{String}(D,N^\prime))}{\log(N/N^\prime)}\end{equation}
The function should collapse into one point at the true critical point $D_\infty$. We simulated the transition for different system sizes $N \in \{32, 40, 50,80,90,100 \}$ using the DMRG algorithm with a maximum bond dimension of $\chi=200$. The data is presented in Fig.~\ref{fig:fssdhaldane}, which shows that the critical point is around $D_\infty = -7.27(1)$.

\subsection{Anistropic-Haldane transition (Line \rom{3})}
  
\begin{figure}[htbp]
 
\includegraphics[width=0.48\textwidth]{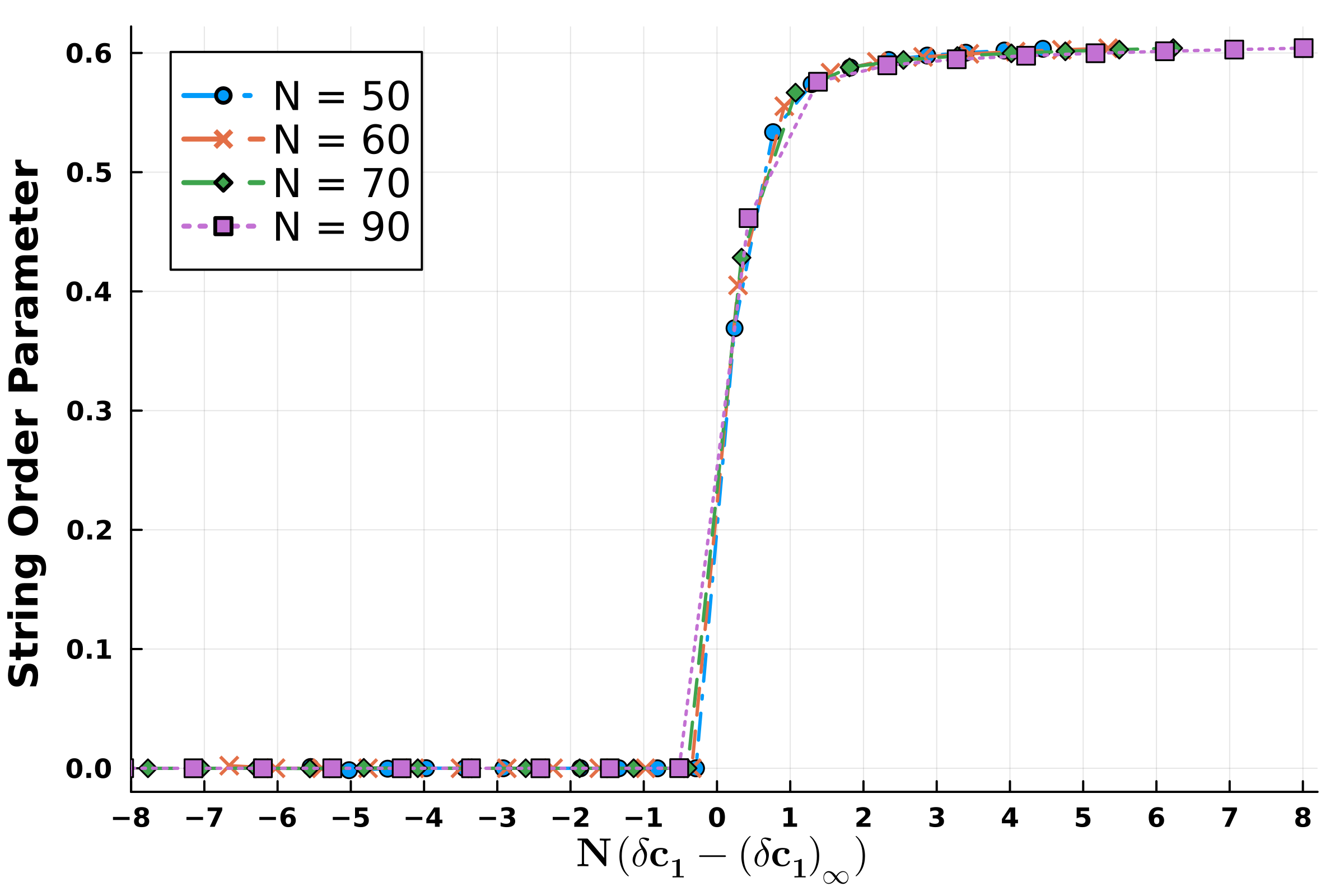}

\caption{\label{fig:fssxyhaldane} String order parameter $ {O}_{String}(\delta c_1, N)$ defined in Eq.~\ref{equation:stringn} as a function of $N(\delta c_1 - (\delta c_1)_{\infty})$ for different system sizes at $D = -6.2$. It shows Line \rom{3} in Fig.~\ref{fig:XY1a} which tracks the transition from the anisotropic to the Haldane phase.  }
\end{figure}

This first-order transition is marked by a very sharp behavior even for relatively small system sizes ($N=26$). We can use the string order parameter to detect this transition since it vanishes in the anisotropic phase. Since the transition is first order, the string order parameter does not show the same power-law dependence expected for continuous phase transitions such as line \rom{2}. We use a data collapse to estimate the true critical point in the form of: 
\begin{equation}\label{equation:stringn}   {O}_{String}(\delta c_1, N) \coloneq  {O}_{String}((\delta c_1-(\delta c_1)_{\infty}) \times N) \end{equation} We simulated the transition for system sizes $N \in \{ 50,60, 70, 90 \}$ using the DMRG algorithm with a maximum bond dimension of $\chi=150$. By fitting $(\delta c_1)_{\infty}$, near perfect collapse is obtained for $(\delta c_1)_{\infty} = -4.48(9)$. The data collapse is shown in Fig.~\ref{fig:fssxyhaldane}.

\subsection{Anisotropic-negative-D transition (Line \rom{4})}
  
\begin{figure}[htbp] 
 
\includegraphics[width=0.48\textwidth]{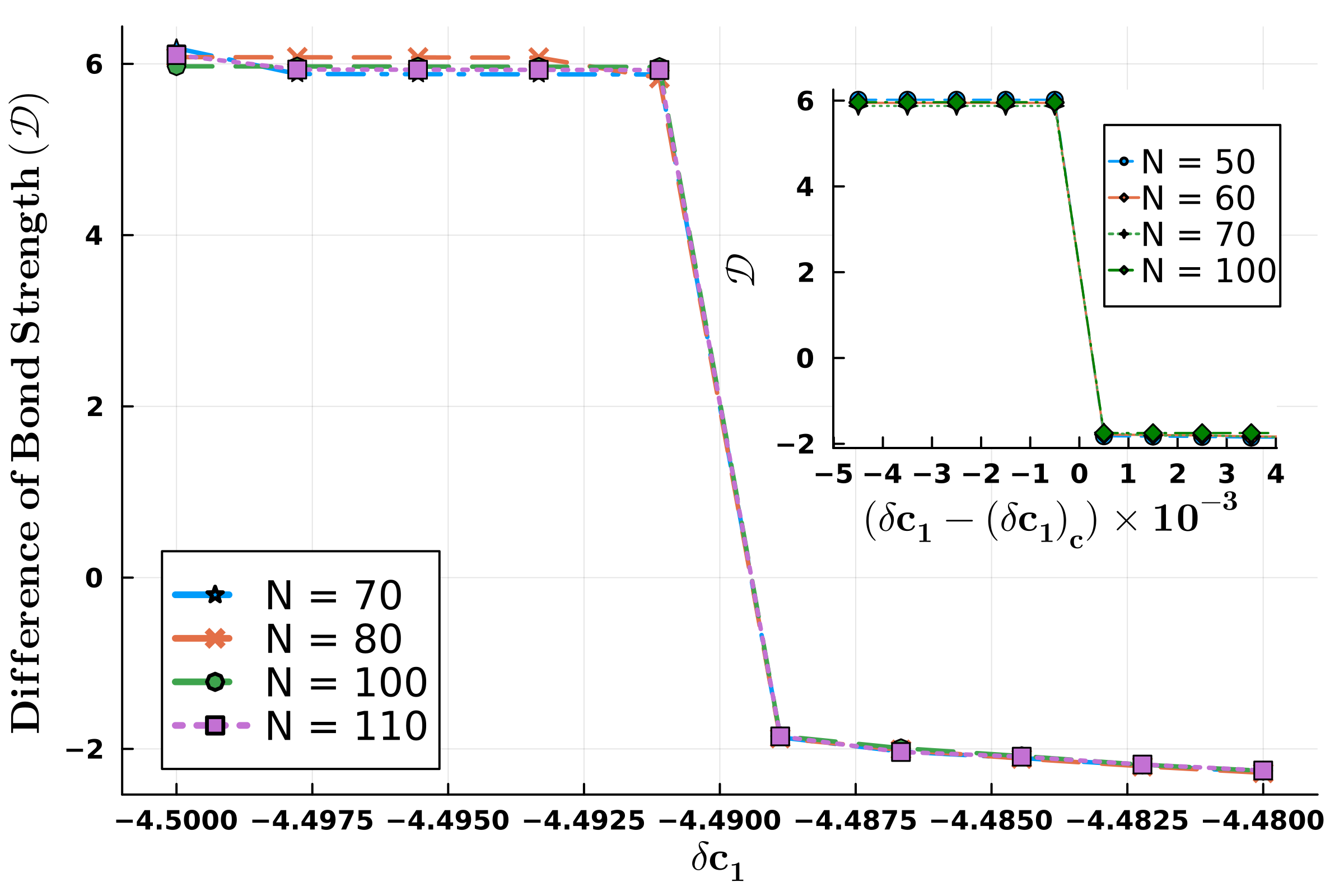}

\caption{\label{fig:fssxynegatived} Difference in bond strength ($\mathcal{D}$) defined in Eq.~\ref{equation:dbs} as a function of $\delta c_1$ for different system sizes at $D = -7.05$. It shows Line \rom{4} in Fig.~\ref{fig:XY1a}. The inset shows the discontinuity of $\mathcal{D}$ around the phase transition with step size = $10^{-3}$. }
\end{figure}

This transition is first order, similar to the anisotropic-Haldane transition (Line \rom{3}). We cannot use the string order parameter to detect the transition as it is zero in both phases, as presented in Table~\ref{tab:ordersandphases}. We can, however, use the ferromagnetic order parameter to detect this transition. Another useful quantity is the difference in bond strength ($\mathcal{D}$), which was proposed to capture first-order quantum phase transitions \cite{Luo2019}. The quantity is related to the level crossing that occurs when we vary a controlling parameter $\lambda$ in the Hamiltonian $H(\lambda) = H_0 + \lambda H_1$, where $H_0$ and $H_1$ do not have to commute. The nonanalyticity of the ground state energy can then be captured using the difference between the expectation value of the two Hamiltonians in the ground state of the system: $\mathcal{D} \coloneq {e}_{0} - \lambda {e}_1$. Where $e_0 \left(e_1\right)$ are the energies of the Hamiltonians $H_0\left(H_1\right)$ per lattice site. In our case, we use:
 
\begin{equation}\label{equation:dbs}
\mathcal{D}(\delta c_1) = \langle H \rangle_{N/2 ,N/2+1} - 2\delta c_1 \langle S^{z}_{N/2} S^{z}_{N/2+1} \rangle,
  \end{equation}
 where $\langle H \rangle_{N/2 ,N/2+1}$ is the energy of the full Hamiltonian Eq.~\ref{equation:ham} evaluated at the bond between the sites $\frac{N}{2}$ and $\frac{N}{2}+1$. The system was simulated using the DMRG algorithm for sizes $N \in \{50,60,70,80,100,110\}$ and a maximum bond dimension of $\chi=200$. The behavior of $\mathcal{D}$ characterizes a first-order transition, as shown in Fig.~\ref{fig:fssxynegatived}. The inset shows that the jump is still evident even when the step size in $\delta c_1$ is of the order $10^{-3}$. The critical point is then estimated at $\delta (c_1)_{\infty} = -4.490(0)$ using this information.

\subsection{Antiferromagentic-Large-D transition (Line \rom{5})}
  
\begin{figure}[htbp]
 
\includegraphics[width=0.48\textwidth]{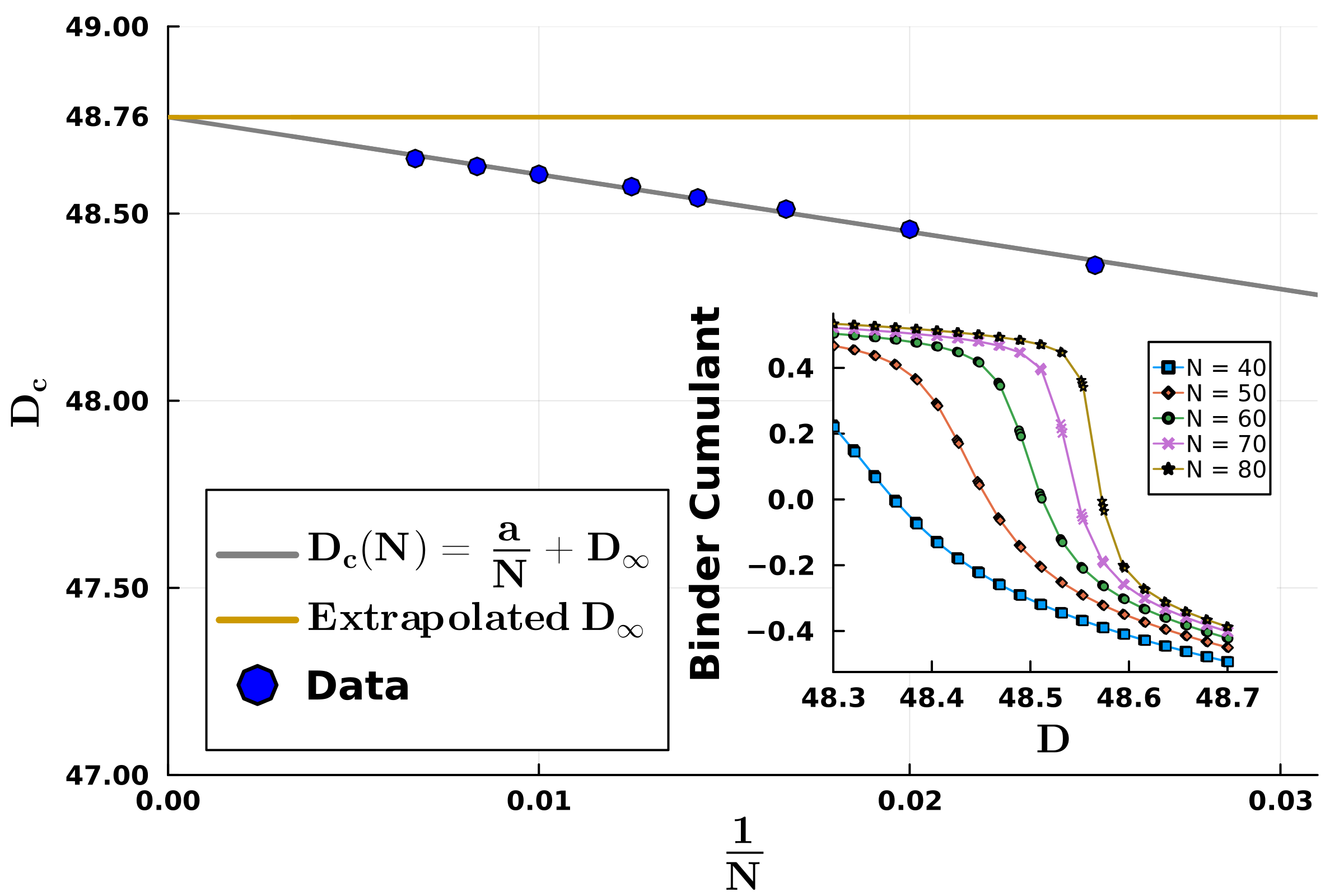}

\caption{\label{fig:fssaftere} Extrapolation of the critical point $D_\infty$ for the transition from the antiferromagnetic phase to the large D phase at $\delta c_1 = 30$ which is Line \rom{5} in Fig.~\ref{fig:phaseb}. A straight line fit for the critical points for finite sizes as a function of  $\frac{1}{N}$ results in a fitted slope of $a = 15.31$. The inset shows the Binder cumulant as a function of $D$ for representative system sizes. }
\end{figure}

We investigate this first-order transition at $\delta c_1 = 30$. The critical points for finite system sizes are obtained and then extrapolated to $N \to \infty$. The Binder cumulant is defined as \cite{Binder1981}:
\begin{equation}
U_N = 1-\frac{\left\langle {S_{z}}^4\right\rangle_N}{3 \left\langle S_z^2\right\rangle_N^2}. \end{equation} The Binder cumulant measures the deviation of the distribution of local observables from the Gaussian distribution. We observe that the Binder cumulant for $S_z$ changes sign from the antiferromagnetic phase to the Large D phase. We can then track the critical point for finite systems by the point where the Binder cumulant crosses zero. We simulated the system for increasing system sizes $N \in \{40, 50, 60, 70, 80, 100, 120, 150 \}$ using the DMRG algorithm with a maximum bond dimension of $\chi = 200$. The critical points for the finite systems were then extrapolated using $D_c(N)= \frac{a}{N} + D_{\infty} $. The data is shown in Fig.~\ref{fig:fssaftere}, where the inset contains a representative subset of the graphs of the Binder cumulant for finite system sizes. The estimated critical point lies at $D_\infty = 48.7(6)$.  
 
\section{Outlook} We constructed and analyzed the rich topological phase diagram for a spin-1 system hosting six different phases and six triple points. The effect of dimerization \cite{Ejima2021-zp} or different trap geometries for Rydberg excitons \cite{Tzeng2017,Ren2018} is expected to exhibit exotic phenomena such as fifth-order transitions \cite{Wu2023}. It would also be interesting to examine these new phase transitions using the photoluminescence spectra of the exciton chain \cite{Poddubny2020} to explore possible applications in the field of quantum simulations.

\section*{Acknowledgments }
The authors are grateful for the internal funding support from the College of Science at Purdue University. S.K. and S.C. would like to acknowledge the support from the Department of Energy, the Office of Science, and the Quantum Science Center (QSC).

    \appendix

\section{Detection of the Haldane Phase}\label{section:appen}
Since Haldane's conjecture in 1983 \cite{Haldane19832}, many properties of the Haldane phase were understood gradually. First of all, the existence of the gap itself can be used to detect the phase when all other phases are gapless. After the AKLT model, we know that the edge states will have spin-$\frac{1}{2}$, which can also be used to detect the phase. A hidden  $\mathbb{Z}_2 \times \mathbb{Z}_2$ symmetry was shown to illustrate how symmetry breaking in one model can be related to the string order in another \cite{HiddenKT1992}. The string order parameter was also developed as a way of detecting the long-range order while having no long-range entanglement \cite{dennijs1989}. The entanglement spectrum of the state, while short-ranged, was shown to have at least double degeneracy \cite{Pollmann2010}. The string order parameter was later shown to work only in the specific case of an abelian symmetry with at least two generators \cite{PollmannGem2012}. The renormalization-group flow of the state is a powerful tool that can also be used to detect and classify nontrivial SPTs \cite{RG2009}. The Haldane phase in spin-1 chains in 1D belongs to the class of symmetry-protected topological phases; the ground state is only topological if a symmetry is imposed on the perturbations of the Hamiltonian. In the absence of symmetry, the ground state can be connected adiabatically to the trivial product state $\ket{0} \otimes \ket{0} \dots$ \cite{Pollmann2012}. It was shown that the Haldane phase can be protected by spatial inversion centered along a bond, time-reversal symmetry, or $\mathbb{Z}_2 \times \mathbb{Z}_2$ internal symmetry \cite{Pollmann2012}. Moreover, qualifying Haldane's original conjecture, it will only be a nontrivial SPT phase in spin chains with odd integer spins \cite{Pollmann2012}. 

\subsection{The Haldane Phase in the AKLT model}
\begin{figure}[htbp]
\centering
\includegraphics[width=0.48\textwidth]{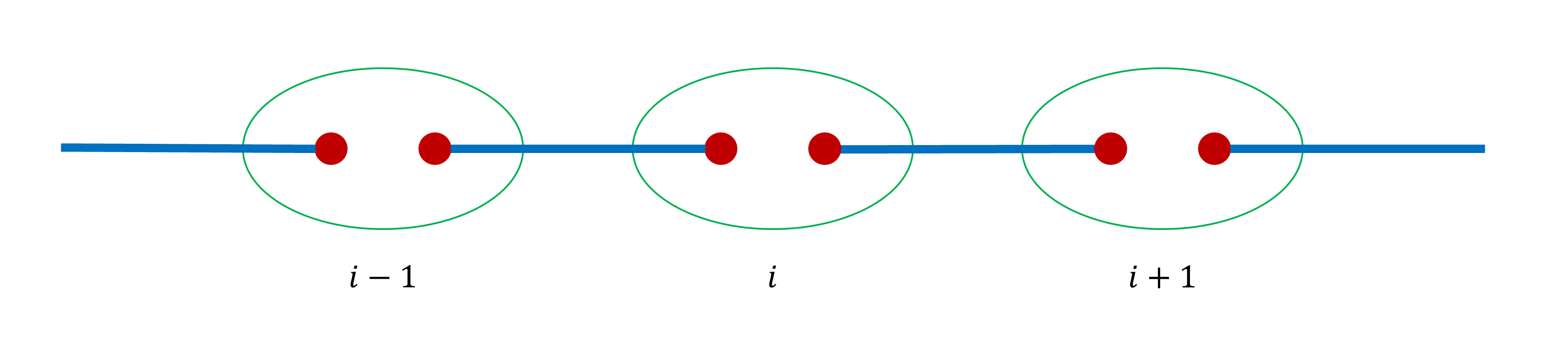} 

\caption{\label{fig:AKLT} Schematic drawing of the ground state of the AKLT model. The red dots represent virtual spin-$\frac{1}{2}$ particles. The green circle designates a symmetrization of the two virtual spins on one site into a physical spin-1 on site $i$. The blue lines represent a singlet state of the two virtual spins on neighboring sites.}
\end{figure}

The AKLT model presented the first exactly solvable realization of the Haldane phase: $$  H_{\mathrm{AKLT}} =\sum_i\left[\frac{1}{2} \mathbf{S}_i \cdot \mathbf{S}_{i+1}+\frac{1}{6}\left(\mathbf{S}_i \cdot \mathbf{S}_{i+1}\right)^2+\frac{1}{3}\right]$$ In the AKLT model, the Hamiltonian is the projection of each two neighboring spin-1 particles into the spin-2 sector:  $H =\sum_i P_2\left(\mathbf{S}_i+\mathbf{S}_{i+1}\right) $. Since the Hamiltonian is a projection, the ground state will have zero energy. The ground state can be formed by ensuring that any two neighboring spins have a total spin of 1 or 0. The method to ensure this is to write the physical spin-1 sites as a symmetric combination of two spin-$\frac{1}{2}$ virtual particles. These spin-$\frac{1}{2}$ particles are then taken to form a singlet bond between neighboring sites, as shown in Fig.~\ref{fig:AKLT}.

 \subsection{Matrix-Product States}\label{mpsconst}
\begin{figure}[htbp]
\centering
\includegraphics[width=0.48\textwidth]{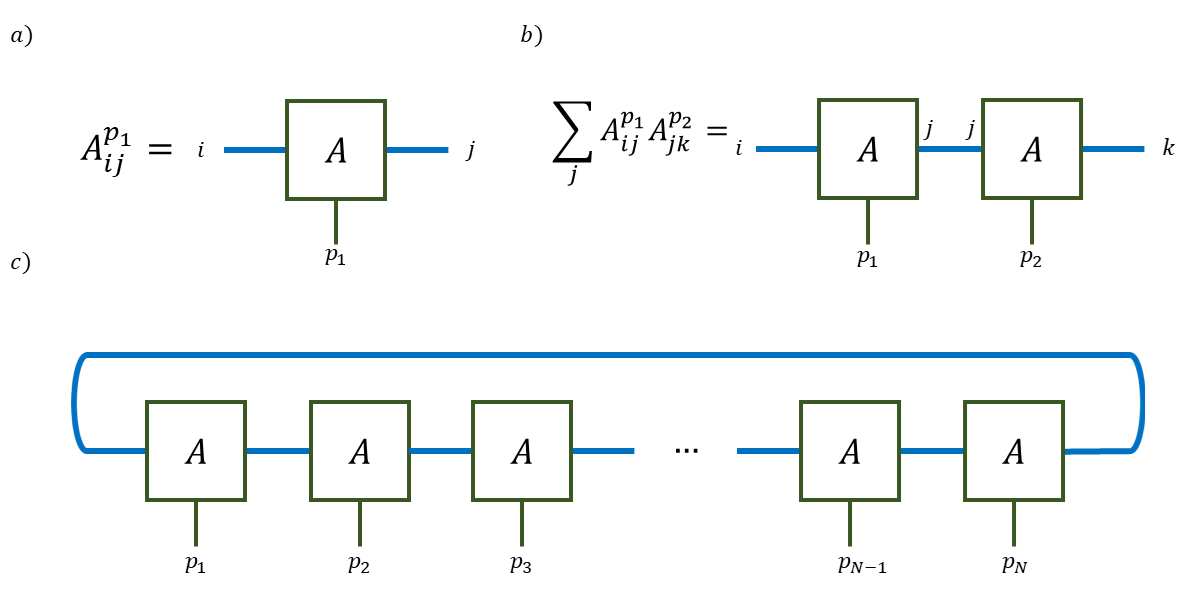} 
\caption{\label{fig:MPS} a) The graphical representation of a rank-three tensor A with the physical index $p_1$ and virtual indices $i,j$. b) Representation of the rank-four tensor $\sum_j A^{p_1}_{i,j}A^{p_2}_{j,k}$, with two physical and two virtual indices, just as Einstein's convention, any connected legs are assumed to be summed over. c) MPS representation of a general translationally invariant state with periodic boundary conditions.}
\end{figure}

Matrix-product states (MPS) play a crucial role in representing 1D systems and offers the theoretical basis for efficient 1D algorithms like DMRG \cite{White1992}. An MPS is a representation of certain wave functions, typically in 1D, as:
\begin{equation}
\ket{\psi} = \Tr ( A^{p_1} A^{p_2} \cdots A^{p_N} ) \ket{p_1} \otimes \ket{p_2} \dots  \otimes \ket{p_N} 
\end{equation}
Here, we represent a translationally invariant periodic state for simplicity. In general, the matrices and their dimensions can be site-dependent. The index $p_i$ runs over the physical degrees of freedom. For a specific $i$, $A^{p_i}$ is a $\chi \times \chi$ matrix. $\chi$ is called the bond dimension or the virtual dimension. The object $A^{p_1}_{\alpha_1 \alpha_2}$ as a whole can be treated as a rank-three tensor. It is often more convenient to use a graphical representation to express MPS operations where tensors are represented by a certain shape and the legs correspond to indices. Linking legs is a convention for summation.  See Fig.~\ref{fig:MPS}.

\begin{figure}[htbp]
\includegraphics[width=0.48\textwidth]{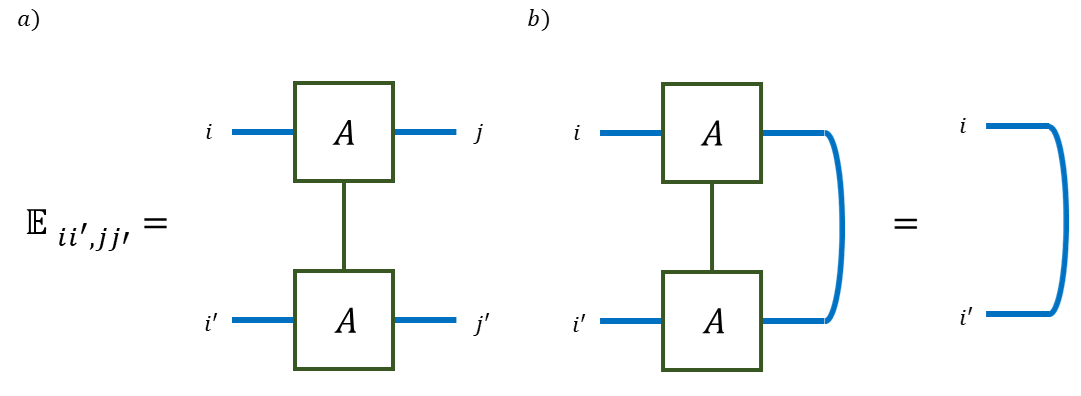} 
\caption{\label{fig:Transfer} a) The graphical representation of the rank-four tensor $\mathbb{E}$ representing the transfer matrix with the implicit summation over the physical index $p$. The matrix A in the bottom row is understood to represent $A^{\dagger}_{i'j'}$. b) The right-canonical condition $\mathbb{E}_{ii',jj'} \delta_{jj'} = \delta_{ii'}  $}
\end{figure}

There is a large gauge freedom $A_{p_1} \to X A_{p_1} X^{-1}$ in the representation of any wavefunction using MPS. This gauge freedom can be exploited to put the matrices in the right-canonical form: $\sum_{p_i}  A^{p_i} A^{\dagger p_i}= \mathbb{I}$ to facilitate both computational and theoretical evaluations of the MPS. If we also define the transfer matrix $\mathbb{E}_{ii',jj'} = \sum_p A^{p}_{ij} A^{\dag p}_{i'j'}$, then the right-canonical form means that $\mathbb{I}$ is a right fixed point for the transfer matrix given in Fig.~\ref{fig:Transfer}. Injective MPS are a finer restriction after this initial gauge fixing. They intuitively represent finitely correlated ground states, which have only short-ranged entanglement. They exclude the so-called cat states which are a superposition of macroscopically different states such as the GHZ state: $\ket{GHZ} = \frac{1}{\sqrt{2}}(\ket{000\cdots} +\ket{111\cdots}$.

The right-canonical form along with injectivity implies the following properties \cite{Fannes1992,Garcia2007}:
\begin{itemize}\label{iteminj}
\item There exists $n_0$ such that $A^{p_1} A^{p_2} \cdots A^{p_n}$ for $(n\geq n_0)$ span the $D\times D$ matrices.
\item $\sum_{p_i}  A^{p_i} A^{ p_i \dagger}= \mathbb{I}$ and for the dual map $\sum_{p_i}  A^{  p_i \dagger} \Lambda A^{p_i} = \Lambda$, where  $\Lambda$ is a full-rank positive diagonal matrix with unit trace.
\item $\mathbb{I}$ is the only right eigenvector with eigenvalue $|\lambda_{\mathbb{I}}|=1$ and all other eigenvalues have strictly smaller magnitudes.
\end{itemize}

\subsection{MPS representation of the AKLT Ground State}
The ground state of the AKLT model has a nice description in terms of three $2 \times 2$ matrices $A^p$, where $p$ stands for the spin index in the  standard basis $S^z=\{-1,0,1\}$ \cite{Fannes1992,Totsuka1995}. 

\begin{equation}
A^{1}=\begin{pmatrix}
0 & 0 \\
-\frac{1}{\sqrt{2}} & 0
\end{pmatrix}, \;
A^0=\begin{pmatrix}
\frac{1}{2} & 0 \\
0 & -\frac{1}{2}
\end{pmatrix}, \;
A^{-1}=\begin{pmatrix}
0 & \frac{1}{\sqrt{2}} \\
0 & 0
\end{pmatrix}
\end{equation}

These matrices can be combined into a rank-three tensor $A^{p_1}_{ij}$; here $p_1$ represents the physical spin index $p_1 \in \{-1,0,1\}$ and $i \text{,} j$ are virtual indices $i,j \in \{1,2\}$. It can be checked that these matrices represent the ground state for open or periodic boundary conditions, respectively: 
\begin{align}
\ket{\Psi_{AKLT}}_{\text{OBC}} &= \sum_{\alpha} \tilde{A}_{\alpha_1}^{p_1} A_{\alpha_1\alpha_2}^{p_2} \cdots \tilde{A}_{\alpha_N}^{p_N} \ket{p_1} \otimes \cdots \otimes \ket{p_N} \\
\ket{\Psi_{AKLT}}_{\text{PBC}} &= \mathrm{Tr}(A^{p_1} A^{p_2} \cdots A^{p_N}) \ket{p_1} \otimes \cdots \otimes \ket{p_N}
\end{align}
Here the summation is over all $\alpha_i$. For open boundary conditions, $\tilde{A}^{p_1}_{\alpha_1}$ is equivalent to ${A}_{\alpha_0 \alpha_1}$, where $\alpha_0$ is a free index representing the unlinked virtual spin-$\frac{1}{2}$ at the left end of the open chain. Similarly, $\tilde{A}^{p_N}_{\alpha_N}$ is equivalent to ${A}_{\alpha_N \alpha_{N+1}}$, with $\alpha_{N+1}$ being a free index.

\subsection{The String Order Parameter in the AKLT Model}
The string order parameter is a non-local order parameter that characterizes the Haldane phase. It was first motivated by the roughening phase transitions of crystals and related to the AKLT model \cite{dennijs1989}:  \begin{align} {O}_{\text{string}} = \lim_{|i-j| \to \infty} \langle S_i^{z}  [ \Pi_{k=i+1}^{j-1} e^{i \pi S_k^{z}}  ] S_j^{z} \rangle \end{align} The configuration of the ground state of the AKLT model amounts to taking the superposition of all states satisfying two rules: there is no restriction on the number or position of sites with spin $S^z = 0$, and no two consecutive sites with spin  $S^z = 1$ or  $S^z = -1$ are allowed (even if separated by sites with  $S^z = 0$). A representative term in the ground state is $\ket{1}\otimes \ket{0}\otimes \ket{0}\otimes \ket{-1}\otimes \ket{1}\otimes \ket{-1}\otimes \ket{1} \otimes \ket{-1}$. This is the usual antiferromagnetic order as realized in the N\'eel state, but with arbitrary zeroes in between, which is called diluted antiferromagnetic order. 

\subsection{Hidden  $\mathbb{Z}_2 \times \mathbb{Z}_2$ Symmetry Breaking}
Another perspective on the Haldane phase and the string order parameter can be seen through a unitary, non-local transformation at each site $j$ \cite{HiddenKT1992,Oshikawa1992}: $U_{\mathrm{KT}}=\prod_{j<k} e^{ i \pi S_j^z S_k^x}$. The transformation is symmetric under the $\mathbb{Z}_2 \times \mathbb{Z}_2$ symmetry, which is represented in our case by the $\pi$-rotations around the three spin axes. The transformation maps the local Hamiltonian (AKLT or Eq.~\eqref{equation:ham}) with open boundary conditions into another short-ranged Hamiltonian. In the new Hamiltonian, the Haldane phase is mapped to a ferromagnetic phase which spontaneously breaks the $\mathbb{Z}_2 \times \mathbb{Z}_2$ symmetry. The ferromagnetic phase is well understood under the Landau symmetry-breaking picture. The correlation function in the infinite limit $\lim_{|i-j| \to \infty} \langle S_z^i S_z^j \rangle$ will have a non-zero value for the ferromagnetic phase. Using the reverse transformation and noting $U_{\mathrm{KT}} U_{\mathrm{KT}} = \mathbb{I}$, we find that this correlation function is mapped to the string order parameter associated with the Haldane phase in the original Hamiltonian. This non-local transformation $U_{\mathrm{KT}}$ then relates the symmetric topological Haldane phase to a ferromagnetic phase, where the symmetry is spontaneously broken.

\subsection{The General String Order Parameter for Abelian Symmetries} \label{subsection:genstring}

\begin{figure}[htbp]
\includegraphics[scale=0.25]{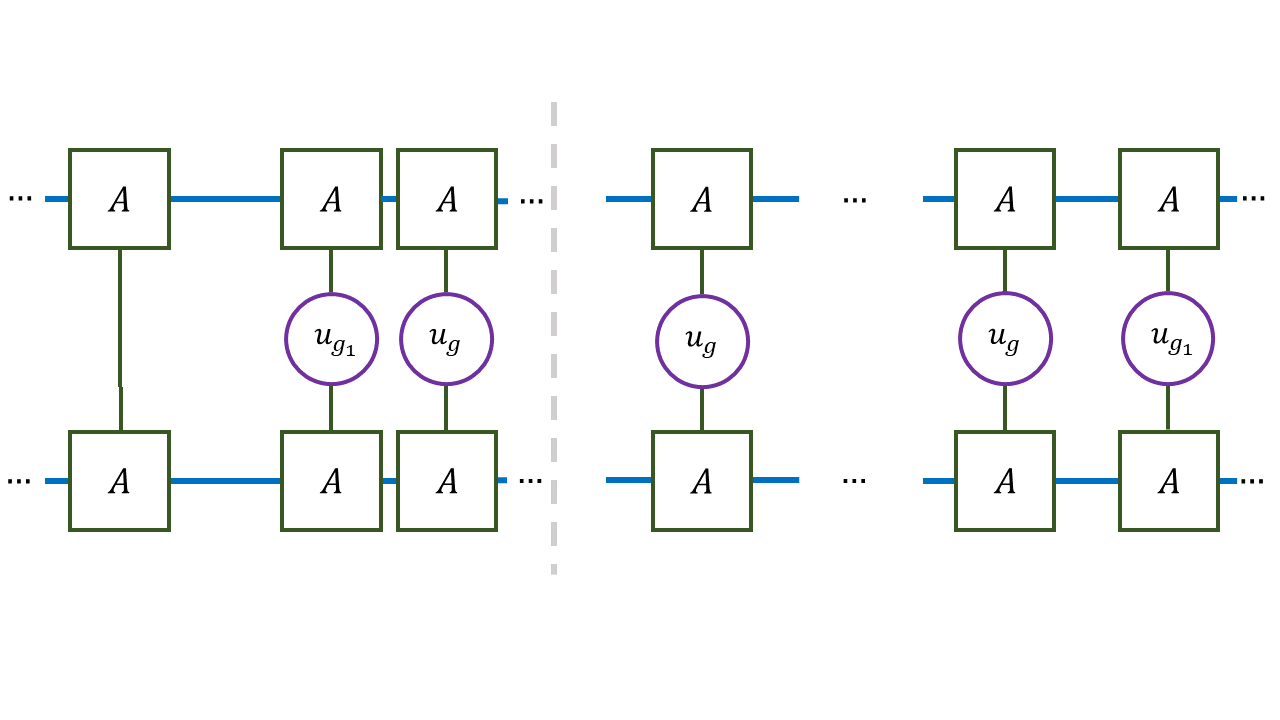} 
\includegraphics[scale=0.25]{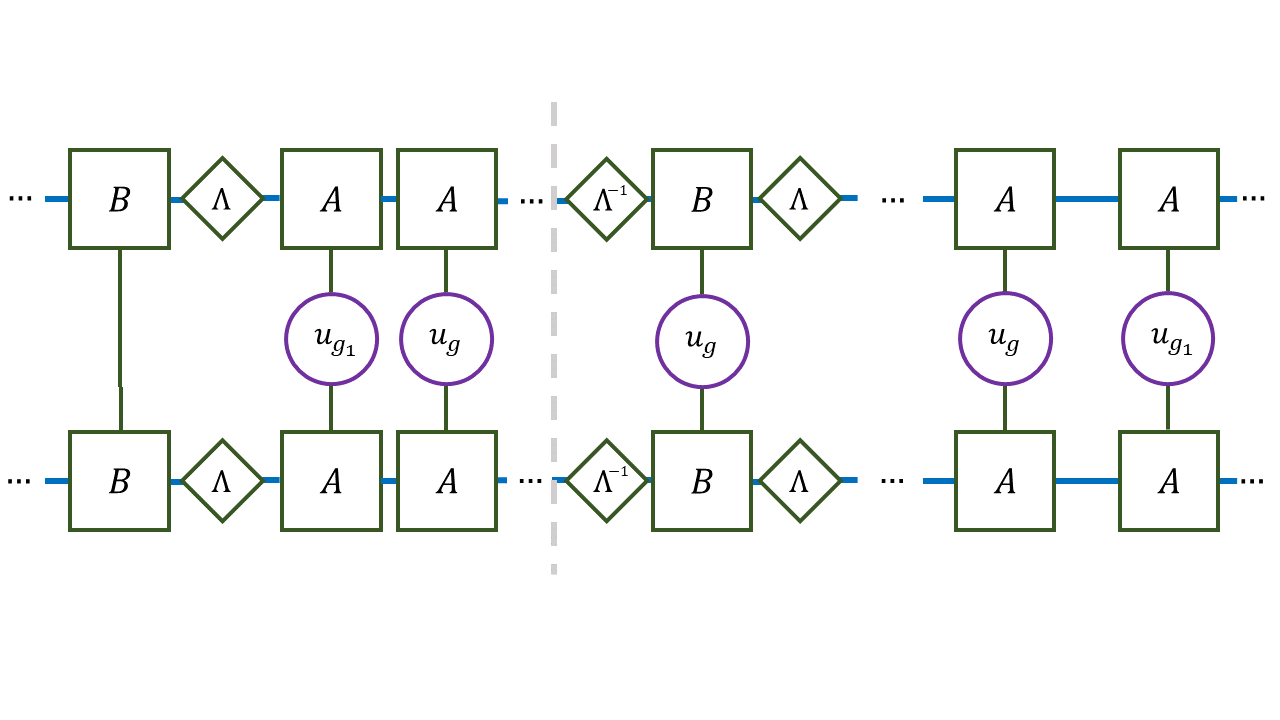}
\includegraphics[scale=0.25]{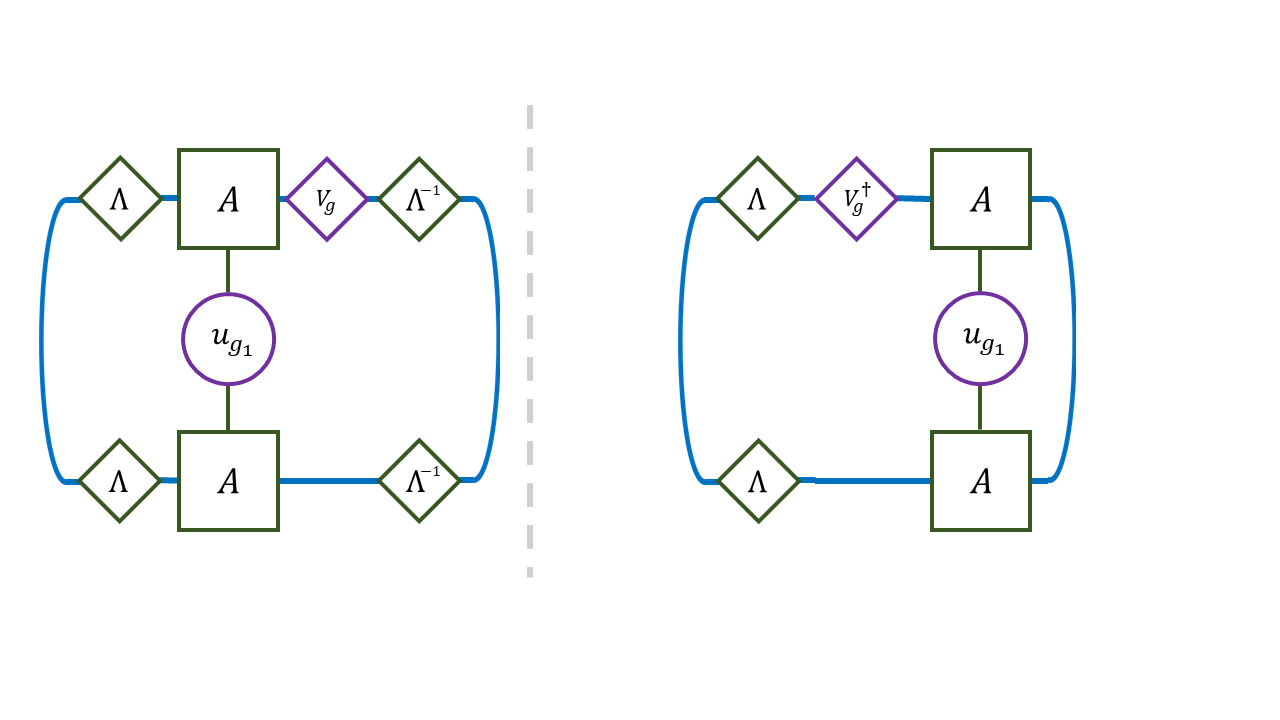}

\caption{\label{fig:strings} Upper: the general string order parameter $\hat{O}_{\text{general string}}$ applied in the physical space. \newline Middle: we can put the left and middle matrices in the left-canonical form $B^{p_i}$. The two forms are related by $A^{p_i} = \Lambda^{-1} B^{p_i} \Lambda $. Lower: for a sufficiently long string, the leftmost and rightmost canonical matrices can be replaced with the identity matrix. The string in the middle can also be seen as acting projectively with unitaries $V_g$ that will cancel except on the edges. }
\end{figure}

In the general setup, we have two symmetries that commute in the physical space, i.e.\ $g, g_2 \in G$ and $[u(g), u(g_2)]=0$, where $G$ is the symmetry group of the Hamiltonian. In the virtual space, however, they only commute  projectively: $V_{g_2} V_{g} = e^{i\phi}  V_{g}V_{g_2} $. Then we can find another operator $g_1$ which  satisfies $u(g_2) u(g_1) = e
^{i\sigma} u(g_1) u(g_2)$. Our general string operator will then be: \begin{equation} \hat{O}_{\text{general string}} = \lim_{N \to \infty} u(g_1)_1  [ \Pi_{j=2}^{N-1} u(g)_j  ] u(g_1)_N \end{equation} In the bulk of the string, we can use injectivity to make $u_g$ act projectively on the virtual dimensions. These unitaries will cancel except on the two edges. If this string is long enough, we can again use injectivity to effectively separate it into two terms. This will give $0$ if $\sigma \neq \phi$ in the two symmetry-respecting phases. In Fig.~\ref{fig:strings} the string order parameter is broken down using the exchange of some right-canonical matrices to left-canonical matrices using $A^{p_i} = \Lambda^{-1} B^{p_i} \Lambda $ where this $\Lambda$ is the same matrix appearing in section \ref{mpsconst}. If the string is long enough, we can exploit the canonical form to reduce the left and right infinite chains to identity \cite{Garcia2008,PollmannGem2012}. We only pick up some $\Lambda$($\Lambda^{-1}$) due to the shift of the orthogonality center. It should be noted that $\Lambda$ will commute with any symmetry operator, as it represents the entanglement spectrum and this should remain invariant under the symmetry.  If we define $L_{{g_1}} = \sum_{ii'}{\Lambda^2_{ii'} \mathbb{E}^{u_{g_1}}_{ii',jj'}} $ and  $R_{{g_1}} = \sum_{j}{ \mathbb{E}^{u_{g_1}}_{ii',jj}} $ for the left and right transfer matrices taking into account the extra $\Lambda$, we arrive at:
\begin{equation}
 {O}_{\text{general string}} = \Tr(L_{g_1} V_g \Lambda^{-2})  \Tr(V_g^{\dagger}R_{{g_1}}\Lambda^{2})
\end{equation}
Since $u(g_2)$ commutes with $u(g_1)$ up to a phase $\sigma$ but, projectively, it commutes with $V_g$ up to another phase $\phi$. The left trace then obeys $ \Tr(L_{g_1} V_g \Lambda^{-2})  =  e^{i (\sigma - \phi)}\Tr(L_{g_1} V_g \Lambda^{-2})  $ and thus it will vanish for $\sigma \neq \phi$ as desired \cite{PollmannGem2012}. In the Haldane phase, we have the elements $g \to e^{i \pi S_x}$, $g_2 \to e^{i \pi S_z}$ and $g_1 \to S_z$. Because the phase is topologically nontrivial, we have $\sigma=\pi$ so $ e^{i \pi S_z}$ and $ e^{i \pi S_x}$ have to anti-commute virtually to give non-zero string order. The two $\pi$-rotations would commute for physical spin-1 edges. However, owing to the topological nature of the phase, the edges are effectively spin-$\frac{1}{2}$ states whose rotation matrices form a representation of SU(2) and anti-commute. 
\newpage
 \bibliography{SPT_MCKKK}

\providecommand{\noopsort}[1]{}\providecommand{\singleletter}[1]{#1}%
\begin{thebibliography}{60}%
\makeatletter
\providecommand \@ifxundefined [1]{%
 \@ifx{#1\undefined}
}%
\providecommand \@ifnum [1]{%
 \ifnum #1\expandafter \@firstoftwo
 \else \expandafter \@secondoftwo
 \fi
}%
\providecommand \@ifx [1]{%
 \ifx #1\expandafter \@firstoftwo
 \else \expandafter \@secondoftwo
 \fi
}%
\providecommand \natexlab [1]{#1}%
\providecommand \enquote  [1]{``#1''}%
\providecommand \bibnamefont  [1]{#1}%
\providecommand \bibfnamefont [1]{#1}%
\providecommand \citenamefont [1]{#1}%
\providecommand \href@noop [0]{\@secondoftwo}%
\providecommand \href [0]{\begingroup \@sanitize@url \@href}%
\providecommand \@href[1]{\@@startlink{#1}\@@href}%
\providecommand \@@href[1]{\endgroup#1\@@endlink}%
\providecommand \@sanitize@url [0]{\catcode `\\12\catcode `\$12\catcode `\&12\catcode `\#12\catcode `\^12\catcode `\_12\catcode `\%12\relax}%
\providecommand \@@startlink[1]{}%
\providecommand \@@endlink[0]{}%
\providecommand \url  [0]{\begingroup\@sanitize@url \@url }%
\providecommand \@url [1]{\endgroup\@href {#1}{\urlprefix }}%
\providecommand \urlprefix  [0]{URL }%
\providecommand \Eprint [0]{\href }%
\providecommand \doibase [0]{https://doi.org/}%
\providecommand \selectlanguage [0]{\@gobble}%
\providecommand \bibinfo  [0]{\@secondoftwo}%
\providecommand \bibfield  [0]{\@secondoftwo}%
\providecommand \translation [1]{[#1]}%
\providecommand \BibitemOpen [0]{}%
\providecommand \bibitemStop [0]{}%
\providecommand \bibitemNoStop [0]{.\EOS\space}%
\providecommand \EOS [0]{\spacefactor3000\relax}%
\providecommand \BibitemShut  [1]{\csname bibitem#1\endcsname}%
\let\auto@bib@innerbib\@empty
\bibitem [{\citenamefont {Kitaev}(2001)}]{Kitaev2001}%
  \BibitemOpen
  \bibfield  {author} {\bibinfo {author} {\bibfnamefont {A.~Y.}\ \bibnamefont {Kitaev}},\ }\bibfield  {title} {\bibinfo {title} {Unpaired majorana fermions in quantum wires},\ }\href {https://doi.org/10.1070/1063-7869/44/10s/s29} {\bibfield  {journal} {\bibinfo  {journal} {Physics-Uspekhi}\ }\textbf {\bibinfo {volume} {44}},\ \bibinfo {pages} {131–136} (\bibinfo {year} {2001})}\BibitemShut {NoStop}%
\bibitem [{\citenamefont {Sarma}\ \emph {et~al.}(2015)\citenamefont {Sarma}, \citenamefont {Freedman},\ and\ \citenamefont {Nayak}}]{Sarma2015}%
  \BibitemOpen
  \bibfield  {author} {\bibinfo {author} {\bibfnamefont {S.~D.}\ \bibnamefont {Sarma}}, \bibinfo {author} {\bibfnamefont {M.}~\bibnamefont {Freedman}},\ and\ \bibinfo {author} {\bibfnamefont {C.}~\bibnamefont {Nayak}},\ }\bibfield  {title} {\bibinfo {title} {Majorana zero modes and topological quantum computation},\ }\href@noop {} {\bibfield  {journal} {\bibinfo  {journal} {Npj Quantum Inf.}\ }\textbf {\bibinfo {volume} {1}} (\bibinfo {year} {2015})}\BibitemShut {NoStop}%
\bibitem [{\citenamefont {Jaworowski}\ and\ \citenamefont {Hawrylak}(2019)}]{Jaworowski2019}%
  \BibitemOpen
  \bibfield  {author} {\bibinfo {author} {\bibfnamefont {B.}~\bibnamefont {Jaworowski}}\ and\ \bibinfo {author} {\bibfnamefont {P.}~\bibnamefont {Hawrylak}},\ }\bibfield  {title} {\bibinfo {title} {Quantum bits with macroscopic topologically protected states in semiconductor devices},\ }\href@noop {} {\bibfield  {journal} {\bibinfo  {journal} {Appl. Sci. (Basel)}\ }\textbf {\bibinfo {volume} {9}},\ \bibinfo {pages} {474} (\bibinfo {year} {2019})}\BibitemShut {NoStop}%
\bibitem [{\citenamefont {Bandres}\ \emph {et~al.}(2018)\citenamefont {Bandres}, \citenamefont {Wittek}, \citenamefont {Harari}, \citenamefont {Parto}, \citenamefont {Ren}, \citenamefont {Segev}, \citenamefont {Christodoulides},\ and\ \citenamefont {Khajavikhan}}]{bandres2018}%
  \BibitemOpen
  \bibfield  {author} {\bibinfo {author} {\bibfnamefont {M.~A.}\ \bibnamefont {Bandres}}, \bibinfo {author} {\bibfnamefont {S.}~\bibnamefont {Wittek}}, \bibinfo {author} {\bibfnamefont {G.}~\bibnamefont {Harari}}, \bibinfo {author} {\bibfnamefont {M.}~\bibnamefont {Parto}}, \bibinfo {author} {\bibfnamefont {J.}~\bibnamefont {Ren}}, \bibinfo {author} {\bibfnamefont {M.}~\bibnamefont {Segev}}, \bibinfo {author} {\bibfnamefont {D.~N.}\ \bibnamefont {Christodoulides}},\ and\ \bibinfo {author} {\bibfnamefont {M.}~\bibnamefont {Khajavikhan}},\ }\bibfield  {title} {\bibinfo {title} {Topological insulator laser: Experiments},\ }\href@noop {} {\bibfield  {journal} {\bibinfo  {journal} {Science}\ }\textbf {\bibinfo {volume} {359}},\ \bibinfo {pages} {eaar4005} (\bibinfo {year} {2018})}\BibitemShut {NoStop}%
\bibitem [{\citenamefont {He}\ \emph {et~al.}(2019)\citenamefont {He}, \citenamefont {Sun},\ and\ \citenamefont {He}}]{He2019}%
  \BibitemOpen
  \bibfield  {author} {\bibinfo {author} {\bibfnamefont {M.}~\bibnamefont {He}}, \bibinfo {author} {\bibfnamefont {H.}~\bibnamefont {Sun}},\ and\ \bibinfo {author} {\bibfnamefont {Q.~L.}\ \bibnamefont {He}},\ }\bibfield  {title} {\bibinfo {title} {Topological insulator: Spintronics and quantum computations},\ }\href@noop {} {\bibfield  {journal} {\bibinfo  {journal} {Front. Phys.}\ }\textbf {\bibinfo {volume} {14}} (\bibinfo {year} {2019})}\BibitemShut {NoStop}%
\bibitem [{\citenamefont {Landau}\ and\ \citenamefont {Lifshitz}(1996)}]{Landau1996}%
  \BibitemOpen
  \bibfield  {author} {\bibinfo {author} {\bibfnamefont {L.~D.}\ \bibnamefont {Landau}}\ and\ \bibinfo {author} {\bibfnamefont {E.~M.}\ \bibnamefont {Lifshitz}},\ }\href@noop {} {\emph {\bibinfo {title} {Statistical Physics}}},\ \bibinfo {edition} {3rd}\ ed.\ (\bibinfo  {publisher} {Butterworth-Heinemann},\ \bibinfo {address} {Oxford, England},\ \bibinfo {year} {1996})\BibitemShut {NoStop}%
\bibitem [{\citenamefont {Klitzing}\ \emph {et~al.}(1980)\citenamefont {Klitzing}, \citenamefont {Dorda},\ and\ \citenamefont {Pepper}}]{hall80}%
  \BibitemOpen
  \bibfield  {author} {\bibinfo {author} {\bibfnamefont {K.~v.}\ \bibnamefont {Klitzing}}, \bibinfo {author} {\bibfnamefont {G.}~\bibnamefont {Dorda}},\ and\ \bibinfo {author} {\bibfnamefont {M.}~\bibnamefont {Pepper}},\ }\bibfield  {title} {\bibinfo {title} {New method for high-accuracy determination of the fine-structure constant based on quantized hall resistance},\ }\href {https://doi.org/10.1103/PhysRevLett.45.494} {\bibfield  {journal} {\bibinfo  {journal} {Phys. Rev. Lett.}\ }\textbf {\bibinfo {volume} {45}},\ \bibinfo {pages} {494} (\bibinfo {year} {1980})}\BibitemShut {NoStop}%
\bibitem [{\citenamefont {Laughlin}(1983)}]{Laughlin1983}%
  \BibitemOpen
  \bibfield  {author} {\bibinfo {author} {\bibfnamefont {R.~B.}\ \bibnamefont {Laughlin}},\ }\bibfield  {title} {\bibinfo {title} {Anomalous quantum hall effect: An incompressible quantum fluid with fractionally charged excitations},\ }\href {https://doi.org/10.1103/PhysRevLett.50.1395} {\bibfield  {journal} {\bibinfo  {journal} {Phys. Rev. Lett.}\ }\textbf {\bibinfo {volume} {50}},\ \bibinfo {pages} {1395} (\bibinfo {year} {1983})}\BibitemShut {NoStop}%
\bibitem [{\citenamefont {Tsui}\ \emph {et~al.}(1982)\citenamefont {Tsui}, \citenamefont {Stormer},\ and\ \citenamefont {Gossard}}]{Tsui1982}%
  \BibitemOpen
  \bibfield  {author} {\bibinfo {author} {\bibfnamefont {D.~C.}\ \bibnamefont {Tsui}}, \bibinfo {author} {\bibfnamefont {H.~L.}\ \bibnamefont {Stormer}},\ and\ \bibinfo {author} {\bibfnamefont {A.~C.}\ \bibnamefont {Gossard}},\ }\bibfield  {title} {\bibinfo {title} {Two-dimensional magnetotransport in the extreme quantum limit},\ }\href@noop {} {\bibfield  {journal} {\bibinfo  {journal} {Phys. Rev. Lett.}\ }\textbf {\bibinfo {volume} {48}},\ \bibinfo {pages} {1559} (\bibinfo {year} {1982})}\BibitemShut {NoStop}%
\bibitem [{\citenamefont {Haldane}(1988)}]{haldane88}%
  \BibitemOpen
  \bibfield  {author} {\bibinfo {author} {\bibfnamefont {F.~D.~M.}\ \bibnamefont {Haldane}},\ }\bibfield  {title} {\bibinfo {title} {Model for a quantum hall effect without landau levels: Condensed-matter realization of the ``parity anomaly''},\ }\href@noop {} {\bibfield  {journal} {\bibinfo  {journal} {Phys. Rev. Lett.}\ }\textbf {\bibinfo {volume} {61}},\ \bibinfo {pages} {2015} (\bibinfo {year} {1988})}\BibitemShut {NoStop}%
\bibitem [{\citenamefont {Haldane}(1983{\natexlab{a}})}]{Haldane1983}%
  \BibitemOpen
  \bibfield  {author} {\bibinfo {author} {\bibfnamefont {F.~D.~M.}\ \bibnamefont {Haldane}},\ }\bibfield  {title} {\bibinfo {title} {Nonlinear field theory of large-spin heisenberg antiferromagnets: Semiclassically quantized solitons of the one-dimensional easy-axis n\'eel state},\ }\href {https://doi.org/10.1103/PhysRevLett.50.1153} {\bibfield  {journal} {\bibinfo  {journal} {Phys. Rev. Lett.}\ }\textbf {\bibinfo {volume} {50}},\ \bibinfo {pages} {1153} (\bibinfo {year} {1983}{\natexlab{a}})}\BibitemShut {NoStop}%
\bibitem [{\citenamefont {Haldane}(1983{\natexlab{b}})}]{Haldane19832}%
  \BibitemOpen
  \bibfield  {author} {\bibinfo {author} {\bibfnamefont {F.~D.~M.}\ \bibnamefont {Haldane}},\ }\bibfield  {title} {\bibinfo {title} {Continuum dynamics of the {1-D} heisenberg antiferromagnet: Identification with the o(3) nonlinear sigma model},\ }\href@noop {} {\bibfield  {journal} {\bibinfo  {journal} {Phys. Lett. A}\ }\textbf {\bibinfo {volume} {93}},\ \bibinfo {pages} {464} (\bibinfo {year} {1983}{\natexlab{b}})}\BibitemShut {NoStop}%
\bibitem [{\citenamefont {Kane}\ and\ \citenamefont {Mele}(2005{\natexlab{a}})}]{Kane2005}%
  \BibitemOpen
  \bibfield  {author} {\bibinfo {author} {\bibfnamefont {C.~L.}\ \bibnamefont {Kane}}\ and\ \bibinfo {author} {\bibfnamefont {E.~J.}\ \bibnamefont {Mele}},\ }\bibfield  {title} {\bibinfo {title} {${Z}_{2}$ topological order and the quantum spin hall effect},\ }\href {https://doi.org/10.1103/PhysRevLett.95.146802} {\bibfield  {journal} {\bibinfo  {journal} {Phys. Rev. Lett.}\ }\textbf {\bibinfo {volume} {95}},\ \bibinfo {pages} {146802} (\bibinfo {year} {2005}{\natexlab{a}})}\BibitemShut {NoStop}%
\bibitem [{\citenamefont {Kane}\ and\ \citenamefont {Mele}(2005{\natexlab{b}})}]{Kane20052}%
  \BibitemOpen
  \bibfield  {author} {\bibinfo {author} {\bibfnamefont {C.~L.}\ \bibnamefont {Kane}}\ and\ \bibinfo {author} {\bibfnamefont {E.~J.}\ \bibnamefont {Mele}},\ }\bibfield  {title} {\bibinfo {title} {Quantum spin hall effect in graphene},\ }\href {https://doi.org/10.1103/PhysRevLett.95.226801} {\bibfield  {journal} {\bibinfo  {journal} {Phys. Rev. Lett.}\ }\textbf {\bibinfo {volume} {95}},\ \bibinfo {pages} {226801} (\bibinfo {year} {2005}{\natexlab{b}})}\BibitemShut {NoStop}%
\bibitem [{\citenamefont {Levin}\ and\ \citenamefont {Wen}(2005)}]{Levin2005}%
  \BibitemOpen
  \bibfield  {author} {\bibinfo {author} {\bibfnamefont {M.~A.}\ \bibnamefont {Levin}}\ and\ \bibinfo {author} {\bibfnamefont {X.-G.}\ \bibnamefont {Wen}},\ }\bibfield  {title} {\bibinfo {title} {String-net condensation: A physical mechanism for topological phases},\ }\href {https://doi.org/10.1103/PhysRevB.71.045110} {\bibfield  {journal} {\bibinfo  {journal} {Phys. Rev. B}\ }\textbf {\bibinfo {volume} {71}},\ \bibinfo {pages} {045110} (\bibinfo {year} {2005})}\BibitemShut {NoStop}%
\bibitem [{\citenamefont {Kitaev}(2003)}]{Kitaev2003}%
  \BibitemOpen
  \bibfield  {author} {\bibinfo {author} {\bibfnamefont {A.~Y.}\ \bibnamefont {Kitaev}},\ }\bibfield  {title} {\bibinfo {title} {Fault-tolerant quantum computation by anyons},\ }\href@noop {} {\bibfield  {journal} {\bibinfo  {journal} {Ann. Phys. (N. Y.)}\ }\textbf {\bibinfo {volume} {303}},\ \bibinfo {pages} {2} (\bibinfo {year} {2003})}\BibitemShut {NoStop}%
\bibitem [{\citenamefont {Wen}(2017)}]{Wen2017}%
  \BibitemOpen
  \bibfield  {author} {\bibinfo {author} {\bibfnamefont {X.-G.}\ \bibnamefont {Wen}},\ }\bibfield  {title} {\bibinfo {title} {Colloquium: Zoo of quantum-topological phases of matter},\ }\href {https://doi.org/10.1103/RevModPhys.89.041004} {\bibfield  {journal} {\bibinfo  {journal} {Rev. Mod. Phys.}\ }\textbf {\bibinfo {volume} {89}},\ \bibinfo {pages} {041004} (\bibinfo {year} {2017})}\BibitemShut {NoStop}%
\bibitem [{\citenamefont {Chen}\ \emph {et~al.}(2010)\citenamefont {Chen}, \citenamefont {Gu},\ and\ \citenamefont {Wen}}]{Chen2010}%
  \BibitemOpen
  \bibfield  {author} {\bibinfo {author} {\bibfnamefont {X.}~\bibnamefont {Chen}}, \bibinfo {author} {\bibfnamefont {Z.-C.}\ \bibnamefont {Gu}},\ and\ \bibinfo {author} {\bibfnamefont {X.-G.}\ \bibnamefont {Wen}},\ }\bibfield  {title} {\bibinfo {title} {Local unitary transformation, long-range quantum entanglement, wave function renormalization, and topological order},\ }\href {https://doi.org/10.1103/PhysRevB.82.155138} {\bibfield  {journal} {\bibinfo  {journal} {Phys. Rev. B}\ }\textbf {\bibinfo {volume} {82}},\ \bibinfo {pages} {155138} (\bibinfo {year} {2010})}\BibitemShut {NoStop}%
\bibitem [{\citenamefont {Mesaros}\ and\ \citenamefont {Ran}(2013)}]{Mesaros2013}%
  \BibitemOpen
  \bibfield  {author} {\bibinfo {author} {\bibfnamefont {A.}~\bibnamefont {Mesaros}}\ and\ \bibinfo {author} {\bibfnamefont {Y.}~\bibnamefont {Ran}},\ }\bibfield  {title} {\bibinfo {title} {Classification of symmetry enriched topological phases with exactly solvable models},\ }\href {https://doi.org/10.1103/PhysRevB.87.155115} {\bibfield  {journal} {\bibinfo  {journal} {Phys. Rev. B}\ }\textbf {\bibinfo {volume} {87}},\ \bibinfo {pages} {155115} (\bibinfo {year} {2013})}\BibitemShut {NoStop}%
\bibitem [{\citenamefont {Chen}\ \emph {et~al.}(2011)\citenamefont {Chen}, \citenamefont {Gu},\ and\ \citenamefont {Wen}}]{Chen2011}%
  \BibitemOpen
  \bibfield  {author} {\bibinfo {author} {\bibfnamefont {X.}~\bibnamefont {Chen}}, \bibinfo {author} {\bibfnamefont {Z.-C.}\ \bibnamefont {Gu}},\ and\ \bibinfo {author} {\bibfnamefont {X.-G.}\ \bibnamefont {Wen}},\ }\bibfield  {title} {\bibinfo {title} {Classification of gapped symmetric phases in one-dimensional spin systems},\ }\href {https://doi.org/10.1103/PhysRevB.83.035107} {\bibfield  {journal} {\bibinfo  {journal} {Phys. Rev. B}\ }\textbf {\bibinfo {volume} {83}},\ \bibinfo {pages} {035107} (\bibinfo {year} {2011})}\BibitemShut {NoStop}%
\bibitem [{\citenamefont {Bethe}(1931)}]{Bethe1931}%
  \BibitemOpen
  \bibfield  {author} {\bibinfo {author} {\bibfnamefont {H.}~\bibnamefont {Bethe}},\ }\bibfield  {title} {\bibinfo {title} {Zur theorie der metalle},\ }\href@noop {} {\bibfield  {journal} {\bibinfo  {journal} {Zeit. für Physik}\ }\textbf {\bibinfo {volume} {71}},\ \bibinfo {pages} {205} (\bibinfo {year} {1931})}\BibitemShut {NoStop}%
\bibitem [{\citenamefont {Affleck}\ \emph {et~al.}(1987)\citenamefont {Affleck}, \citenamefont {Kennedy}, \citenamefont {Lieb},\ and\ \citenamefont {Tasaki}}]{AKLT1987}%
  \BibitemOpen
  \bibfield  {author} {\bibinfo {author} {\bibfnamefont {I.}~\bibnamefont {Affleck}}, \bibinfo {author} {\bibfnamefont {T.}~\bibnamefont {Kennedy}}, \bibinfo {author} {\bibfnamefont {E.~H.}\ \bibnamefont {Lieb}},\ and\ \bibinfo {author} {\bibfnamefont {H.}~\bibnamefont {Tasaki}},\ }\bibfield  {title} {\bibinfo {title} {Rigorous results on valence-bond ground states in antiferromagnets},\ }\href {https://doi.org/10.1103/PhysRevLett.59.799} {\bibfield  {journal} {\bibinfo  {journal} {Phys. Rev. Lett.}\ }\textbf {\bibinfo {volume} {59}},\ \bibinfo {pages} {799} (\bibinfo {year} {1987})}\BibitemShut {NoStop}%
\bibitem [{\citenamefont {Pollmann}\ \emph {et~al.}(2012)\citenamefont {Pollmann}, \citenamefont {Berg}, \citenamefont {Turner},\ and\ \citenamefont {Oshikawa}}]{Pollmann2012}%
  \BibitemOpen
  \bibfield  {author} {\bibinfo {author} {\bibfnamefont {F.}~\bibnamefont {Pollmann}}, \bibinfo {author} {\bibfnamefont {E.}~\bibnamefont {Berg}}, \bibinfo {author} {\bibfnamefont {A.~M.}\ \bibnamefont {Turner}},\ and\ \bibinfo {author} {\bibfnamefont {M.}~\bibnamefont {Oshikawa}},\ }\bibfield  {title} {\bibinfo {title} {Symmetry protection of topological phases in one-dimensional quantum spin systems},\ }\href {https://doi.org/10.1103/PhysRevB.85.075125} {\bibfield  {journal} {\bibinfo  {journal} {Phys. Rev. B}\ }\textbf {\bibinfo {volume} {85}},\ \bibinfo {pages} {075125} (\bibinfo {year} {2012})}\BibitemShut {NoStop}%
\bibitem [{\citenamefont {Kl{\"u}mper}\ \emph {et~al.}(1993)\citenamefont {Kl{\"u}mper}, \citenamefont {Schadschneider},\ and\ \citenamefont {Zittartz}}]{Klumper1993}%
  \BibitemOpen
  \bibfield  {author} {\bibinfo {author} {\bibfnamefont {A.}~\bibnamefont {Kl{\"u}mper}}, \bibinfo {author} {\bibfnamefont {A.}~\bibnamefont {Schadschneider}},\ and\ \bibinfo {author} {\bibfnamefont {J.}~\bibnamefont {Zittartz}},\ }\bibfield  {title} {\bibinfo {title} {Matrix product ground states for one-dimensional spin-1 quantum antiferromagnets},\ }\href@noop {} {\bibfield  {journal} {\bibinfo  {journal} {EPL}\ }\textbf {\bibinfo {volume} {24}},\ \bibinfo {pages} {293} (\bibinfo {year} {1993})}\BibitemShut {NoStop}%
\bibitem [{\citenamefont {Walther}\ \emph {et~al.}(2018)\citenamefont {Walther}, \citenamefont {Kr\"uger}, \citenamefont {Scheel},\ and\ \citenamefont {Pohl}}]{Valentin2018}%
  \BibitemOpen
  \bibfield  {author} {\bibinfo {author} {\bibfnamefont {V.}~\bibnamefont {Walther}}, \bibinfo {author} {\bibfnamefont {S.~O.}\ \bibnamefont {Kr\"uger}}, \bibinfo {author} {\bibfnamefont {S.}~\bibnamefont {Scheel}},\ and\ \bibinfo {author} {\bibfnamefont {T.}~\bibnamefont {Pohl}},\ }\bibfield  {title} {\bibinfo {title} {Interactions between rydberg excitons in ${\mathrm{cu}}_{2}\mathrm{O}$},\ }\href {https://doi.org/10.1103/PhysRevB.98.165201} {\bibfield  {journal} {\bibinfo  {journal} {Phys. Rev. B}\ }\textbf {\bibinfo {volume} {98}},\ \bibinfo {pages} {165201} (\bibinfo {year} {2018})}\BibitemShut {NoStop}%
\bibitem [{\citenamefont {Poddubny}\ and\ \citenamefont {Glazov}(2019)}]{Poddubny2019}%
  \BibitemOpen
  \bibfield  {author} {\bibinfo {author} {\bibfnamefont {A.~N.}\ \bibnamefont {Poddubny}}\ and\ \bibinfo {author} {\bibfnamefont {M.~M.}\ \bibnamefont {Glazov}},\ }\bibfield  {title} {\bibinfo {title} {Topological spin phases of trapped rydberg excitons in ${\mathrm{cu}}_{2}\mathrm{O}$},\ }\href {https://doi.org/10.1103/PhysRevLett.123.126801} {\bibfield  {journal} {\bibinfo  {journal} {Phys. Rev. Lett.}\ }\textbf {\bibinfo {volume} {123}},\ \bibinfo {pages} {126801} (\bibinfo {year} {2019})}\BibitemShut {NoStop}%
\bibitem [{\citenamefont {Urban}\ \emph {et~al.}(2009)\citenamefont {Urban}, \citenamefont {Johnson}, \citenamefont {Henage}, \citenamefont {Isenhower}, \citenamefont {Yavuz}, \citenamefont {Walker},\ and\ \citenamefont {Saffman}}]{Urban2009}%
  \BibitemOpen
  \bibfield  {author} {\bibinfo {author} {\bibfnamefont {E.}~\bibnamefont {Urban}}, \bibinfo {author} {\bibfnamefont {T.~A.}\ \bibnamefont {Johnson}}, \bibinfo {author} {\bibfnamefont {T.}~\bibnamefont {Henage}}, \bibinfo {author} {\bibfnamefont {L.}~\bibnamefont {Isenhower}}, \bibinfo {author} {\bibfnamefont {D.~D.}\ \bibnamefont {Yavuz}}, \bibinfo {author} {\bibfnamefont {T.~G.}\ \bibnamefont {Walker}},\ and\ \bibinfo {author} {\bibfnamefont {M.}~\bibnamefont {Saffman}},\ }\bibfield  {title} {\bibinfo {title} {Observation of rydberg blockade between two atoms},\ }\href@noop {} {\bibfield  {journal} {\bibinfo  {journal} {Nat. Phys.}\ }\textbf {\bibinfo {volume} {5}},\ \bibinfo {pages} {110} (\bibinfo {year} {2009})}\BibitemShut {NoStop}%
\bibitem [{\citenamefont {Knox}(1963)}]{Knox1963}%
  \BibitemOpen
  \bibfield  {author} {\bibinfo {author} {\bibfnamefont {R.~S.}\ \bibnamefont {Knox}},\ }\href@noop {} {\emph {\bibinfo {title} {Theory of Excitons}}}\ (\bibinfo  {publisher} {Academic Press},\ \bibinfo {address} {San Diego, CA},\ \bibinfo {year} {1963})\BibitemShut {NoStop}%
\bibitem [{\citenamefont {Peyronel}\ \emph {et~al.}(2012)\citenamefont {Peyronel}, \citenamefont {Firstenberg}, \citenamefont {Liang}, \citenamefont {Hofferberth}, \citenamefont {Gorshkov}, \citenamefont {Pohl}, \citenamefont {Lukin},\ and\ \citenamefont {Vuleti{\'c}}}]{Peyronel2012-no}%
  \BibitemOpen
  \bibfield  {author} {\bibinfo {author} {\bibfnamefont {T.}~\bibnamefont {Peyronel}}, \bibinfo {author} {\bibfnamefont {O.}~\bibnamefont {Firstenberg}}, \bibinfo {author} {\bibfnamefont {Q.-Y.}\ \bibnamefont {Liang}}, \bibinfo {author} {\bibfnamefont {S.}~\bibnamefont {Hofferberth}}, \bibinfo {author} {\bibfnamefont {A.~V.}\ \bibnamefont {Gorshkov}}, \bibinfo {author} {\bibfnamefont {T.}~\bibnamefont {Pohl}}, \bibinfo {author} {\bibfnamefont {M.~D.}\ \bibnamefont {Lukin}},\ and\ \bibinfo {author} {\bibfnamefont {V.}~\bibnamefont {Vuleti{\'c}}},\ }\bibfield  {title} {\bibinfo {title} {Quantum nonlinear optics with single photons enabled by strongly interacting atoms},\ }\href@noop {} {\bibfield  {journal} {\bibinfo  {journal} {Nature}\ }\textbf {\bibinfo {volume} {488}},\ \bibinfo {pages} {57} (\bibinfo {year} {2012})}\BibitemShut {NoStop}%
\bibitem [{\citenamefont {Bernien}\ \emph {et~al.}(2017)\citenamefont {Bernien}, \citenamefont {Schwartz}, \citenamefont {Keesling}, \citenamefont {Levine}, \citenamefont {Omran}, \citenamefont {Pichler}, \citenamefont {Choi}, \citenamefont {Zibrov}, \citenamefont {Endres}, \citenamefont {Greiner}, \citenamefont {Vuleti{\'c}},\ and\ \citenamefont {Lukin}}]{Bernien2017}%
  \BibitemOpen
  \bibfield  {author} {\bibinfo {author} {\bibfnamefont {H.}~\bibnamefont {Bernien}}, \bibinfo {author} {\bibfnamefont {S.}~\bibnamefont {Schwartz}}, \bibinfo {author} {\bibfnamefont {A.}~\bibnamefont {Keesling}}, \bibinfo {author} {\bibfnamefont {H.}~\bibnamefont {Levine}}, \bibinfo {author} {\bibfnamefont {A.}~\bibnamefont {Omran}}, \bibinfo {author} {\bibfnamefont {H.}~\bibnamefont {Pichler}}, \bibinfo {author} {\bibfnamefont {S.}~\bibnamefont {Choi}}, \bibinfo {author} {\bibfnamefont {A.~S.}\ \bibnamefont {Zibrov}}, \bibinfo {author} {\bibfnamefont {M.}~\bibnamefont {Endres}}, \bibinfo {author} {\bibfnamefont {M.}~\bibnamefont {Greiner}}, \bibinfo {author} {\bibfnamefont {V.}~\bibnamefont {Vuleti{\'c}}},\ and\ \bibinfo {author} {\bibfnamefont {M.~D.}\ \bibnamefont {Lukin}},\ }\bibfield  {title} {\bibinfo {title} {Probing many-body dynamics on a 51-atom quantum simulator},\ }\href@noop {} {\bibfield  {journal} {\bibinfo  {journal} {Nature}\ }\textbf {\bibinfo {volume} {551}},\ \bibinfo {pages} {579}
  (\bibinfo {year} {2017})}\BibitemShut {NoStop}%
\bibitem [{\citenamefont {Bluvstein}\ \emph {et~al.}(2023)\citenamefont {Bluvstein}, \citenamefont {Evered}, \citenamefont {Geim}, \citenamefont {Li}, \citenamefont {Zhou}, \citenamefont {Manovitz}, \citenamefont {Ebadi}, \citenamefont {Cain}, \citenamefont {Kalinowski}, \citenamefont {Hangleiter}, \citenamefont {Ataides}, \citenamefont {Maskara}, \citenamefont {Cong}, \citenamefont {Gao}, \citenamefont {Rodriguez}, \citenamefont {Karolyshyn}, \citenamefont {Semeghini}, \citenamefont {Gullans}, \citenamefont {Greiner}, \citenamefont {Vuleti{\'c}},\ and\ \citenamefont {Lukin}}]{Bluvstein2023-gr}%
  \BibitemOpen
  \bibfield  {author} {\bibinfo {author} {\bibfnamefont {D.}~\bibnamefont {Bluvstein}}, \bibinfo {author} {\bibfnamefont {S.~J.}\ \bibnamefont {Evered}}, \bibinfo {author} {\bibfnamefont {A.~A.}\ \bibnamefont {Geim}}, \bibinfo {author} {\bibfnamefont {S.~H.}\ \bibnamefont {Li}}, \bibinfo {author} {\bibfnamefont {H.}~\bibnamefont {Zhou}}, \bibinfo {author} {\bibfnamefont {T.}~\bibnamefont {Manovitz}}, \bibinfo {author} {\bibfnamefont {S.}~\bibnamefont {Ebadi}}, \bibinfo {author} {\bibfnamefont {M.}~\bibnamefont {Cain}}, \bibinfo {author} {\bibfnamefont {M.}~\bibnamefont {Kalinowski}}, \bibinfo {author} {\bibfnamefont {D.}~\bibnamefont {Hangleiter}}, \bibinfo {author} {\bibfnamefont {J.~P.~B.}\ \bibnamefont {Ataides}}, \bibinfo {author} {\bibfnamefont {N.}~\bibnamefont {Maskara}}, \bibinfo {author} {\bibfnamefont {I.}~\bibnamefont {Cong}}, \bibinfo {author} {\bibfnamefont {X.}~\bibnamefont {Gao}}, \bibinfo {author} {\bibfnamefont {P.~S.}\ \bibnamefont {Rodriguez}}, \bibinfo {author} {\bibfnamefont
  {T.}~\bibnamefont {Karolyshyn}}, \bibinfo {author} {\bibfnamefont {G.}~\bibnamefont {Semeghini}}, \bibinfo {author} {\bibfnamefont {M.~J.}\ \bibnamefont {Gullans}}, \bibinfo {author} {\bibfnamefont {M.}~\bibnamefont {Greiner}}, \bibinfo {author} {\bibfnamefont {V.}~\bibnamefont {Vuleti{\'c}}},\ and\ \bibinfo {author} {\bibfnamefont {M.~D.}\ \bibnamefont {Lukin}},\ }\bibfield  {title} {\bibinfo {title} {Logical quantum processor based on reconfigurable atom arrays},\ }\href@noop {} {\bibfield  {journal} {\bibinfo  {journal} {Nature}\ } (\bibinfo {year} {2023})}\BibitemShut {NoStop}%
\bibitem [{\citenamefont {Taylor}\ \emph {et~al.}(2022)\citenamefont {Taylor}, \citenamefont {Goswami}, \citenamefont {Walther}, \citenamefont {Spanner}, \citenamefont {Simon},\ and\ \citenamefont {Heshami}}]{Taylor2022}%
  \BibitemOpen
  \bibfield  {author} {\bibinfo {author} {\bibfnamefont {J.}~\bibnamefont {Taylor}}, \bibinfo {author} {\bibfnamefont {S.}~\bibnamefont {Goswami}}, \bibinfo {author} {\bibfnamefont {V.}~\bibnamefont {Walther}}, \bibinfo {author} {\bibfnamefont {M.}~\bibnamefont {Spanner}}, \bibinfo {author} {\bibfnamefont {C.}~\bibnamefont {Simon}},\ and\ \bibinfo {author} {\bibfnamefont {K.}~\bibnamefont {Heshami}},\ }\bibfield  {title} {\bibinfo {title} {Simulation of many-body dynamics using rydberg excitons},\ }\href@noop {} {\bibfield  {journal} {\bibinfo  {journal} {Quantum Sci. Technol.}\ }\textbf {\bibinfo {volume} {7}},\ \bibinfo {pages} {035016} (\bibinfo {year} {2022})}\BibitemShut {NoStop}%
\bibitem [{\citenamefont {Sajjan}\ \emph {et~al.}(2022)\citenamefont {Sajjan}, \citenamefont {Alaeian},\ and\ \citenamefont {Kais}}]{sajjan2022}%
  \BibitemOpen
  \bibfield  {author} {\bibinfo {author} {\bibfnamefont {M.}~\bibnamefont {Sajjan}}, \bibinfo {author} {\bibfnamefont {H.}~\bibnamefont {Alaeian}},\ and\ \bibinfo {author} {\bibfnamefont {S.}~\bibnamefont {Kais}},\ }\bibfield  {title} {\bibinfo {title} {Magnetic phases of spatially modulated spin-1 chains in rydberg excitons: Classical and quantum simulations},\ }\bibfield  {journal} {\bibinfo  {journal} {The Journal of Chemical Physics}\ }\textbf {\bibinfo {volume} {157}},\ \href {https://doi.org/10.1063/5.0128283} {10.1063/5.0128283} (\bibinfo {year} {2022})\BibitemShut {NoStop}%
\bibitem [{\citenamefont {Singer}\ \emph {et~al.}(2005)\citenamefont {Singer}, \citenamefont {Stanojevic}, \citenamefont {Weidem\"{u}ller},\ and\ \citenamefont {C\^oté}}]{Singer2005}%
  \BibitemOpen
  \bibfield  {author} {\bibinfo {author} {\bibfnamefont {K.}~\bibnamefont {Singer}}, \bibinfo {author} {\bibfnamefont {J.}~\bibnamefont {Stanojevic}}, \bibinfo {author} {\bibfnamefont {M.}~\bibnamefont {Weidem\"{u}ller}},\ and\ \bibinfo {author} {\bibfnamefont {R.}~\bibnamefont {C\^oté}},\ }\bibfield  {title} {\bibinfo {title} {Long-range interactions between alkali rydberg atom pairs correlated to thens–ns, np–np andnd–nd asymptotes},\ }\href {https://doi.org/10.1088/0953-4075/38/2/021} {\bibfield  {journal} {\bibinfo  {journal} {Journal of Physics B: Atomic, Molecular and Optical Physics}\ }\textbf {\bibinfo {volume} {38}},\ \bibinfo {pages} {S295–S307} (\bibinfo {year} {2005})}\BibitemShut {NoStop}%
\bibitem [{\citenamefont {den Nijs}\ and\ \citenamefont {Rommelse}(1989)}]{dennijs1989}%
  \BibitemOpen
  \bibfield  {author} {\bibinfo {author} {\bibfnamefont {M.}~\bibnamefont {den Nijs}}\ and\ \bibinfo {author} {\bibfnamefont {K.}~\bibnamefont {Rommelse}},\ }\bibfield  {title} {\bibinfo {title} {Preroughening transitions in crystal surfaces and valence-bond phases in quantum spin chains},\ }\href {https://doi.org/10.1103/PhysRevB.40.4709} {\bibfield  {journal} {\bibinfo  {journal} {Phys. Rev. B}\ }\textbf {\bibinfo {volume} {40}},\ \bibinfo {pages} {4709} (\bibinfo {year} {1989})}\BibitemShut {NoStop}%
\bibitem [{\citenamefont {White}(1992)}]{White1992}%
  \BibitemOpen
  \bibfield  {author} {\bibinfo {author} {\bibfnamefont {S.~R.}\ \bibnamefont {White}},\ }\bibfield  {title} {\bibinfo {title} {Density matrix formulation for quantum renormalization groups},\ }\href {https://doi.org/10.1103/PhysRevLett.69.2863} {\bibfield  {journal} {\bibinfo  {journal} {Phys. Rev. Lett.}\ }\textbf {\bibinfo {volume} {69}},\ \bibinfo {pages} {2863} (\bibinfo {year} {1992})}\BibitemShut {NoStop}%
\bibitem [{\citenamefont {Fishman}\ \emph {et~al.}(2022)\citenamefont {Fishman}, \citenamefont {White},\ and\ \citenamefont {Stoudenmire}}]{itensor}%
  \BibitemOpen
  \bibfield  {author} {\bibinfo {author} {\bibfnamefont {M.}~\bibnamefont {Fishman}}, \bibinfo {author} {\bibfnamefont {S.~R.}\ \bibnamefont {White}},\ and\ \bibinfo {author} {\bibfnamefont {E.~M.}\ \bibnamefont {Stoudenmire}},\ }\bibfield  {title} {\bibinfo {title} {{The ITensor Software Library for Tensor Network Calculations}},\ }\href {https://doi.org/10.21468/SciPostPhysCodeb.4} {\bibfield  {journal} {\bibinfo  {journal} {SciPost Phys. Codebases}\ ,\ \bibinfo {pages} {4}} (\bibinfo {year} {2022})}\BibitemShut {NoStop}%
\bibitem [{\citenamefont {Zauner-Stauber}\ \emph {et~al.}(2018)\citenamefont {Zauner-Stauber}, \citenamefont {Vanderstraeten}, \citenamefont {Fishman}, \citenamefont {Verstraete},\ and\ \citenamefont {Haegeman}}]{VUMPS}%
  \BibitemOpen
  \bibfield  {author} {\bibinfo {author} {\bibfnamefont {V.}~\bibnamefont {Zauner-Stauber}}, \bibinfo {author} {\bibfnamefont {L.}~\bibnamefont {Vanderstraeten}}, \bibinfo {author} {\bibfnamefont {M.~T.}\ \bibnamefont {Fishman}}, \bibinfo {author} {\bibfnamefont {F.}~\bibnamefont {Verstraete}},\ and\ \bibinfo {author} {\bibfnamefont {J.}~\bibnamefont {Haegeman}},\ }\bibfield  {title} {\bibinfo {title} {Variational optimization algorithms for uniform matrix product states},\ }\href {https://doi.org/10.1103/PhysRevB.97.045145} {\bibfield  {journal} {\bibinfo  {journal} {Phys. Rev. B}\ }\textbf {\bibinfo {volume} {97}},\ \bibinfo {pages} {045145} (\bibinfo {year} {2018})}\BibitemShut {NoStop}%
\bibitem [{\citenamefont {Fannes}\ \emph {et~al.}(1992)\citenamefont {Fannes}, \citenamefont {Nachtergaele},\ and\ \citenamefont {Werner}}]{Fannes1992}%
  \BibitemOpen
  \bibfield  {author} {\bibinfo {author} {\bibfnamefont {M.}~\bibnamefont {Fannes}}, \bibinfo {author} {\bibfnamefont {B.}~\bibnamefont {Nachtergaele}},\ and\ \bibinfo {author} {\bibfnamefont {R.~F.}\ \bibnamefont {Werner}},\ }\bibfield  {title} {\bibinfo {title} {Finitely correlated states on quantum spin chains},\ }\href@noop {} {\bibfield  {journal} {\bibinfo  {journal} {Commun. Math. Phys.}\ }\textbf {\bibinfo {volume} {144}},\ \bibinfo {pages} {443} (\bibinfo {year} {1992})}\BibitemShut {NoStop}%
\bibitem [{\citenamefont {Perez-Garcia}\ \emph {et~al.}(2007)\citenamefont {Perez-Garcia}, \citenamefont {Verstraete}, \citenamefont {Wolf},\ and\ \citenamefont {Cirac}}]{Garcia2007}%
  \BibitemOpen
  \bibfield  {author} {\bibinfo {author} {\bibfnamefont {D.}~\bibnamefont {Perez-Garcia}}, \bibinfo {author} {\bibfnamefont {F.}~\bibnamefont {Verstraete}}, \bibinfo {author} {\bibfnamefont {M.~M.}\ \bibnamefont {Wolf}},\ and\ \bibinfo {author} {\bibfnamefont {J.~I.}\ \bibnamefont {Cirac}},\ }\href@noop {} {\bibinfo {title} {Matrix product state representations}} (\bibinfo {year} {2007}),\ \Eprint {https://arxiv.org/abs/quant-ph/0608197} {arXiv:quant-ph/0608197 [quant-ph]} \BibitemShut {NoStop}%
\bibitem [{\citenamefont {Fisher}\ and\ \citenamefont {Barber}(1972)}]{Fisher1972}%
  \BibitemOpen
  \bibfield  {author} {\bibinfo {author} {\bibfnamefont {M.~E.}\ \bibnamefont {Fisher}}\ and\ \bibinfo {author} {\bibfnamefont {M.~N.}\ \bibnamefont {Barber}},\ }\bibfield  {title} {\bibinfo {title} {Scaling theory for finite-size effects in the critical region},\ }\href {https://doi.org/10.1103/PhysRevLett.28.1516} {\bibfield  {journal} {\bibinfo  {journal} {Phys. Rev. Lett.}\ }\textbf {\bibinfo {volume} {28}},\ \bibinfo {pages} {1516} (\bibinfo {year} {1972})}\BibitemShut {NoStop}%
\bibitem [{\citenamefont {Ren}\ \emph {et~al.}(2020)\citenamefont {Ren}, \citenamefont {You},\ and\ \citenamefont {Ole\ifmmode~\acute{s}\else \'{s}\fi{}}}]{Ren2020}%
  \BibitemOpen
  \bibfield  {author} {\bibinfo {author} {\bibfnamefont {J.}~\bibnamefont {Ren}}, \bibinfo {author} {\bibfnamefont {W.-L.}\ \bibnamefont {You}},\ and\ \bibinfo {author} {\bibfnamefont {A.~M.}\ \bibnamefont {Ole\ifmmode~\acute{s}\else \'{s}\fi{}}},\ }\bibfield  {title} {\bibinfo {title} {Quantum phase transitions in a spin-1 antiferromagnetic chain with long-range interactions and modulated single-ion anisotropy},\ }\href {https://doi.org/10.1103/PhysRevB.102.024425} {\bibfield  {journal} {\bibinfo  {journal} {Phys. Rev. B}\ }\textbf {\bibinfo {volume} {102}},\ \bibinfo {pages} {024425} (\bibinfo {year} {2020})}\BibitemShut {NoStop}%
\bibitem [{\citenamefont {Gu}(2010)}]{GU2010}%
  \BibitemOpen
  \bibfield  {author} {\bibinfo {author} {\bibfnamefont {S.-J.}\ \bibnamefont {Gu}},\ }\bibfield  {title} {\bibinfo {title} {Fidelity approach to quantum phase transitions},\ }\href {https://doi.org/10.1142/s0217979210056335} {\bibfield  {journal} {\bibinfo  {journal} {International Journal of Modern Physics B}\ }\textbf {\bibinfo {volume} {24}},\ \bibinfo {pages} {4371–4458} (\bibinfo {year} {2010})}\BibitemShut {NoStop}%
\bibitem [{\citenamefont {Kais}\ and\ \citenamefont {Serra}(2003)}]{Kais2003}%
  \BibitemOpen
  \bibfield  {author} {\bibinfo {author} {\bibfnamefont {S.}~\bibnamefont {Kais}}\ and\ \bibinfo {author} {\bibfnamefont {P.}~\bibnamefont {Serra}},\ }\bibfield  {title} {\bibinfo {title} {Finite‐size scaling for atomic and molecular systems},\ }\href {https://doi.org/10.1002/0471428027.ch1} {\bibfield  {journal} {\bibinfo  {journal} {Advances in Chemical Physics}\ ,\ \bibinfo {pages} {1–99}} (\bibinfo {year} {2003})}\BibitemShut {NoStop}%
\bibitem [{\citenamefont {Tonooka}\ \emph {et~al.}(2007)\citenamefont {Tonooka}, \citenamefont {Nakano}, \citenamefont {Kusakabe},\ and\ \citenamefont {Suzuki}}]{Tonooka2007}%
  \BibitemOpen
  \bibfield  {author} {\bibinfo {author} {\bibfnamefont {S.}~\bibnamefont {Tonooka}}, \bibinfo {author} {\bibfnamefont {H.}~\bibnamefont {Nakano}}, \bibinfo {author} {\bibfnamefont {K.}~\bibnamefont {Kusakabe}},\ and\ \bibinfo {author} {\bibfnamefont {N.}~\bibnamefont {Suzuki}},\ }\bibfield  {title} {\bibinfo {title} {Extended string order parameter in the (s=1, 1/2) mixed spin chain},\ }\href {https://doi.org/10.1143/jpsj.76.084714} {\bibfield  {journal} {\bibinfo  {journal} {Journal of the Physical Society of Japan}\ }\textbf {\bibinfo {volume} {76}},\ \bibinfo {pages} {084714} (\bibinfo {year} {2007})}\BibitemShut {NoStop}%
\bibitem [{\citenamefont {Ueda}\ \emph {et~al.}(2008)\citenamefont {Ueda}, \citenamefont {Nakano},\ and\ \citenamefont {Kusakabe}}]{Ueda2008}%
  \BibitemOpen
  \bibfield  {author} {\bibinfo {author} {\bibfnamefont {H.}~\bibnamefont {Ueda}}, \bibinfo {author} {\bibfnamefont {H.}~\bibnamefont {Nakano}},\ and\ \bibinfo {author} {\bibfnamefont {K.}~\bibnamefont {Kusakabe}},\ }\bibfield  {title} {\bibinfo {title} {Finite-size scaling of string order parameters characterizing the haldane phase},\ }\href {https://doi.org/10.1103/PhysRevB.78.224402} {\bibfield  {journal} {\bibinfo  {journal} {Phys. Rev. B}\ }\textbf {\bibinfo {volume} {78}},\ \bibinfo {pages} {224402} (\bibinfo {year} {2008})}\BibitemShut {NoStop}%
\bibitem [{\citenamefont {Luo}\ \emph {et~al.}(2019)\citenamefont {Luo}, \citenamefont {Zhao},\ and\ \citenamefont {Wang}}]{Luo2019}%
  \BibitemOpen
  \bibfield  {author} {\bibinfo {author} {\bibfnamefont {Q.}~\bibnamefont {Luo}}, \bibinfo {author} {\bibfnamefont {J.}~\bibnamefont {Zhao}},\ and\ \bibinfo {author} {\bibfnamefont {X.}~\bibnamefont {Wang}},\ }\bibfield  {title} {\bibinfo {title} {Intrinsic jump character of first-order quantum phase transitions},\ }\href {https://doi.org/10.1103/PhysRevB.100.121111} {\bibfield  {journal} {\bibinfo  {journal} {Phys. Rev. B}\ }\textbf {\bibinfo {volume} {100}},\ \bibinfo {pages} {121111} (\bibinfo {year} {2019})}\BibitemShut {NoStop}%
\bibitem [{\citenamefont {Binder}(1981)}]{Binder1981}%
  \BibitemOpen
  \bibfield  {author} {\bibinfo {author} {\bibfnamefont {K.}~\bibnamefont {Binder}},\ }\bibfield  {title} {\bibinfo {title} {Finite size scaling analysis of ising model block distribution functions},\ }\href {https://doi.org/10.1007/bf01293604} {\bibfield  {journal} {\bibinfo  {journal} {Zeitschrift f{\"u}r Physik B Condensed Matter}\ }\textbf {\bibinfo {volume} {43}},\ \bibinfo {pages} {119–140} (\bibinfo {year} {1981})}\BibitemShut {NoStop}%
\bibitem [{\citenamefont {Ejima}\ \emph {et~al.}(2021)\citenamefont {Ejima}, \citenamefont {Lange},\ and\ \citenamefont {Fehske}}]{Ejima2021-zp}%
  \BibitemOpen
  \bibfield  {author} {\bibinfo {author} {\bibfnamefont {S.}~\bibnamefont {Ejima}}, \bibinfo {author} {\bibfnamefont {F.}~\bibnamefont {Lange}},\ and\ \bibinfo {author} {\bibfnamefont {H.}~\bibnamefont {Fehske}},\ }\bibfield  {title} {\bibinfo {title} {Quantum criticality in dimerised anisotropic spin-1 chains},\ }\href@noop {} {\bibfield  {journal} {\bibinfo  {journal} {Eur. Phys. J. Spec. Top.}\ }\textbf {\bibinfo {volume} {230}},\ \bibinfo {pages} {1009} (\bibinfo {year} {2021})}\BibitemShut {NoStop}%
\bibitem [{\citenamefont {Tzeng}\ \emph {et~al.}(2017)\citenamefont {Tzeng}, \citenamefont {Onishi}, \citenamefont {Okubo},\ and\ \citenamefont {Kao}}]{Tzeng2017}%
  \BibitemOpen
  \bibfield  {author} {\bibinfo {author} {\bibfnamefont {Y.-C.}\ \bibnamefont {Tzeng}}, \bibinfo {author} {\bibfnamefont {H.}~\bibnamefont {Onishi}}, \bibinfo {author} {\bibfnamefont {T.}~\bibnamefont {Okubo}},\ and\ \bibinfo {author} {\bibfnamefont {Y.-J.}\ \bibnamefont {Kao}},\ }\bibfield  {title} {\bibinfo {title} {Quantum phase transitions driven by rhombic-type single-ion anisotropy in the $s=1$ haldane chain},\ }\href {https://doi.org/10.1103/PhysRevB.96.060404} {\bibfield  {journal} {\bibinfo  {journal} {Phys. Rev. B}\ }\textbf {\bibinfo {volume} {96}},\ \bibinfo {pages} {060404} (\bibinfo {year} {2017})}\BibitemShut {NoStop}%
\bibitem [{\citenamefont {Ren}\ \emph {et~al.}(2018)\citenamefont {Ren}, \citenamefont {Wang},\ and\ \citenamefont {You}}]{Ren2018}%
  \BibitemOpen
  \bibfield  {author} {\bibinfo {author} {\bibfnamefont {J.}~\bibnamefont {Ren}}, \bibinfo {author} {\bibfnamefont {Y.}~\bibnamefont {Wang}},\ and\ \bibinfo {author} {\bibfnamefont {W.-L.}\ \bibnamefont {You}},\ }\bibfield  {title} {\bibinfo {title} {Quantum phase transitions in spin-1 $xxz$ chains with rhombic single-ion anisotropy},\ }\href {https://doi.org/10.1103/PhysRevA.97.042318} {\bibfield  {journal} {\bibinfo  {journal} {Phys. Rev. A}\ }\textbf {\bibinfo {volume} {97}},\ \bibinfo {pages} {042318} (\bibinfo {year} {2018})}\BibitemShut {NoStop}%
\bibitem [{\citenamefont {Wu}\ \emph {et~al.}(2023)\citenamefont {Wu}, \citenamefont {Tzeng}, \citenamefont {Xie}, \citenamefont {Ji},\ and\ \citenamefont {Yu}}]{Wu2023}%
  \BibitemOpen
  \bibfield  {author} {\bibinfo {author} {\bibfnamefont {H.~Y.}\ \bibnamefont {Wu}}, \bibinfo {author} {\bibfnamefont {Y.-C.}\ \bibnamefont {Tzeng}}, \bibinfo {author} {\bibfnamefont {Z.~Y.}\ \bibnamefont {Xie}}, \bibinfo {author} {\bibfnamefont {K.}~\bibnamefont {Ji}},\ and\ \bibinfo {author} {\bibfnamefont {J.~F.}\ \bibnamefont {Yu}},\ }\bibfield  {title} {\bibinfo {title} {Exploring quantum phase transitions by the cross derivative of the ground state energy},\ }\href@noop {} {\bibfield  {journal} {\bibinfo  {journal} {New J. Phys.}\ }\textbf {\bibinfo {volume} {25}},\ \bibinfo {pages} {043006} (\bibinfo {year} {2023})}\BibitemShut {NoStop}%
\bibitem [{\citenamefont {Poddubny}\ and\ \citenamefont {Glazov}(2020)}]{Poddubny2020}%
  \BibitemOpen
  \bibfield  {author} {\bibinfo {author} {\bibfnamefont {A.~N.}\ \bibnamefont {Poddubny}}\ and\ \bibinfo {author} {\bibfnamefont {M.~M.}\ \bibnamefont {Glazov}},\ }\bibfield  {title} {\bibinfo {title} {Polarized edge state emission from topological spin phases of trapped rydberg excitons in ${\mathrm{cu}}_{2}\mathrm{O}$},\ }\href {https://doi.org/10.1103/PhysRevB.102.125307} {\bibfield  {journal} {\bibinfo  {journal} {Phys. Rev. B}\ }\textbf {\bibinfo {volume} {102}},\ \bibinfo {pages} {125307} (\bibinfo {year} {2020})}\BibitemShut {NoStop}%
\bibitem [{\citenamefont {Kennedy}\ and\ \citenamefont {Tasaki}(1992)}]{HiddenKT1992}%
  \BibitemOpen
  \bibfield  {author} {\bibinfo {author} {\bibfnamefont {T.}~\bibnamefont {Kennedy}}\ and\ \bibinfo {author} {\bibfnamefont {H.}~\bibnamefont {Tasaki}},\ }\bibfield  {title} {\bibinfo {title} {Hidden {$Z_2 \times Z_2$ symmetry} breaking in haldane-gap antiferromagnets},\ }\href {https://doi.org/10.1103/PhysRevB.45.304} {\bibfield  {journal} {\bibinfo  {journal} {Phys. Rev. B}\ }\textbf {\bibinfo {volume} {45}},\ \bibinfo {pages} {304} (\bibinfo {year} {1992})}\BibitemShut {NoStop}%
\bibitem [{\citenamefont {Pollmann}\ \emph {et~al.}(2010)\citenamefont {Pollmann}, \citenamefont {Turner}, \citenamefont {Berg},\ and\ \citenamefont {Oshikawa}}]{Pollmann2010}%
  \BibitemOpen
  \bibfield  {author} {\bibinfo {author} {\bibfnamefont {F.}~\bibnamefont {Pollmann}}, \bibinfo {author} {\bibfnamefont {A.~M.}\ \bibnamefont {Turner}}, \bibinfo {author} {\bibfnamefont {E.}~\bibnamefont {Berg}},\ and\ \bibinfo {author} {\bibfnamefont {M.}~\bibnamefont {Oshikawa}},\ }\bibfield  {title} {\bibinfo {title} {Entanglement spectrum of a topological phase in one dimension},\ }\href {https://doi.org/10.1103/PhysRevB.81.064439} {\bibfield  {journal} {\bibinfo  {journal} {Phys. Rev. B}\ }\textbf {\bibinfo {volume} {81}},\ \bibinfo {pages} {064439} (\bibinfo {year} {2010})}\BibitemShut {NoStop}%
\bibitem [{\citenamefont {Pollmann}\ and\ \citenamefont {Turner}(2012)}]{PollmannGem2012}%
  \BibitemOpen
  \bibfield  {author} {\bibinfo {author} {\bibfnamefont {F.}~\bibnamefont {Pollmann}}\ and\ \bibinfo {author} {\bibfnamefont {A.~M.}\ \bibnamefont {Turner}},\ }\bibfield  {title} {\bibinfo {title} {Detection of symmetry-protected topological phases in one dimension},\ }\href {https://doi.org/10.1103/PhysRevB.86.125441} {\bibfield  {journal} {\bibinfo  {journal} {Phys. Rev. B}\ }\textbf {\bibinfo {volume} {86}},\ \bibinfo {pages} {125441} (\bibinfo {year} {2012})}\BibitemShut {NoStop}%
\bibitem [{\citenamefont {Gu}\ and\ \citenamefont {Wen}(2009)}]{RG2009}%
  \BibitemOpen
  \bibfield  {author} {\bibinfo {author} {\bibfnamefont {Z.-C.}\ \bibnamefont {Gu}}\ and\ \bibinfo {author} {\bibfnamefont {X.-G.}\ \bibnamefont {Wen}},\ }\bibfield  {title} {\bibinfo {title} {Tensor-entanglement-filtering renormalization approach and symmetry-protected topological order},\ }\href {https://doi.org/10.1103/PhysRevB.80.155131} {\bibfield  {journal} {\bibinfo  {journal} {Phys. Rev. B}\ }\textbf {\bibinfo {volume} {80}},\ \bibinfo {pages} {155131} (\bibinfo {year} {2009})}\BibitemShut {NoStop}%
\bibitem [{\citenamefont {Totsuka}\ and\ \citenamefont {Suzuki}(1995)}]{Totsuka1995}%
  \BibitemOpen
  \bibfield  {author} {\bibinfo {author} {\bibfnamefont {K.}~\bibnamefont {Totsuka}}\ and\ \bibinfo {author} {\bibfnamefont {M.}~\bibnamefont {Suzuki}},\ }\bibfield  {title} {\bibinfo {title} {Matrix formalism for the {VBS-type} models and hidden order},\ }\href@noop {} {\bibfield  {journal} {\bibinfo  {journal} {J. Phys. Condens. Matter}\ }\textbf {\bibinfo {volume} {7}},\ \bibinfo {pages} {1639} (\bibinfo {year} {1995})}\BibitemShut {NoStop}%
\bibitem [{\citenamefont {Oshikawa}(1992)}]{Oshikawa1992}%
  \BibitemOpen
  \bibfield  {author} {\bibinfo {author} {\bibfnamefont {M.}~\bibnamefont {Oshikawa}},\ }\bibfield  {title} {\bibinfo {title} {Hidden {$Z_2 \times Z_2$ symmetry} in quantum spin chains with arbitrary integer spin},\ }\href@noop {} {\bibfield  {journal} {\bibinfo  {journal} {J. Phys. Condens. Matter}\ }\textbf {\bibinfo {volume} {4}},\ \bibinfo {pages} {7469} (\bibinfo {year} {1992})}\BibitemShut {NoStop}%
\bibitem [{\citenamefont {P\'erez-Garc\'{\i}a}\ \emph {et~al.}(2008)\citenamefont {P\'erez-Garc\'{\i}a}, \citenamefont {Wolf}, \citenamefont {Sanz}, \citenamefont {Verstraete},\ and\ \citenamefont {Cirac}}]{Garcia2008}%
  \BibitemOpen
  \bibfield  {author} {\bibinfo {author} {\bibfnamefont {D.}~\bibnamefont {P\'erez-Garc\'{\i}a}}, \bibinfo {author} {\bibfnamefont {M.~M.}\ \bibnamefont {Wolf}}, \bibinfo {author} {\bibfnamefont {M.}~\bibnamefont {Sanz}}, \bibinfo {author} {\bibfnamefont {F.}~\bibnamefont {Verstraete}},\ and\ \bibinfo {author} {\bibfnamefont {J.~I.}\ \bibnamefont {Cirac}},\ }\bibfield  {title} {\bibinfo {title} {String order and symmetries in quantum spin lattices},\ }\href {https://doi.org/10.1103/PhysRevLett.100.167202} {\bibfield  {journal} {\bibinfo  {journal} {Phys. Rev. Lett.}\ }\textbf {\bibinfo {volume} {100}},\ \bibinfo {pages} {167202} (\bibinfo {year} {2008})}\BibitemShut {NoStop}%
\end{thebibliography}%

\end{document}